\documentclass{elsarticle}

\usepackage{xcolor}

\usepackage{booktabs}

\usepackage{mathtools}

\usepackage{amssymb}

\usepackage{bm}

\usepackage[T1]{fontenc}
\usepackage{lmodern}

\usepackage{tikz}
\usepackage{pgfplots}
\pgfplotsset{compat=1.15}
\pgfplotsset{every tick label/.append style={font=\small}}

\usepackage{hyperref}

\usepackage{tikz-cd}

\usepackage{caption}
\usepackage{subcaption}

\usepackage[ruled]{algorithm}
\usepackage{algorithmic}

\usepackage{gensymb}

\usepackage{placeins}

\usepackage[%
    activate={true,nocompatibility},%
    final,tracking=true,%
    kerning=true,%
    spacing=true,%
    factor=1100,%
    stretch=10,%
shrink=10]{microtype}

\microtypecontext{spacing=nonfrench}

\DeclareMathOperator*{\argmin}{arg\,min} 
\makeatletter 
\let\oldtheequation\theequation
\renewcommand\tagform@[1]{\maketag@@@{\ignorespaces#1\unskip\@@italiccorr}}
\renewcommand\theequation{(\oldtheequation)} 
\makeatother

\def\rightfanD{\mathrel{\hbox{\rlap{$-$}$<$}}}
\let\rightfanT\rightfanD
\def\rightfanS{\mathrel{\hbox{\rlap{$\scriptstyle-$}$\scriptstyle<$}}}
\def\rightfanSS{%
\mathrel{\hbox{\rlap{$\scriptscriptstyle-$}$\scriptscriptstyle<$}}}

\let\crowr\rightfan     

\newcommand{\ah}[1]{\textcolor{black}{#1}}

\begin{document}
\begin{frontmatter}
    \title{Three-dimensional microstructure generation using generative
    adversarial neural networks in the context of continuum micromechanics}
    \author{Alexander Henkes\corref{cor1}
    \fnref{label1, label2}}
    \ead{a.henkes@tu-braunschweig.de}
    \cortext[cor1]{Corresponding author}
    \fntext[label2]{https://orcid.org/0000-0003-4615-9271}
    \author{Henning Wessels%
    \fnref{label1}}
    \affiliation[label1]{
        organization={Institute for Computational Modeling in Civil Engineering,
        Technical University of Braunschweig},
        addressline={Pockelsstr. 3}, 
        city={Braunschweig},
        postcode={38106}, 
    country={Germany}}

    \begin{abstract}
        Multiscale simulations are demanding in terms of  computational
        resources. In the context of continuum micromechanics, the multiscale
        problem arises from the need of inferring macroscopic material
        parameters from the microscale.  If the underlying microstructure is
        explicitly given by means of $\mu$CT-scans, convolutional neural
        networks can be used to learn the microstructure-property mapping, which
        is usually obtained from computational homogenization. The CNN approach
        provides a significant speedup, especially in the context of
        heterogeneous or functionally graded materials. Another application is
        uncertainty quantification, where many expansive evaluations are
        required. However, one bottleneck of this approach is the large number
        of training microstructures needed.

        This work closes this gap by proposing a generative adversarial network
        tailored towards three-dimensional microstructure generation. The
        lightweight algorithm is able to learn the underlying properties of the
        material from a single $\mu$CT-scan without the need of explicit
        descriptors. During prediction time, the network
        can produce unique three-dimensional microstructures with the same
        properties of the original data in a fraction of seconds and at
        consistently high quality.
    \end{abstract}


    \setcounter{tocdepth}{3}

\end{frontmatter}

\section{Introduction} \noindent
Modern composite materials in aerospace, automotive, civil and mechanical
engineering consist of multiple constituents, which induce macroscopic
properties by means of their microstructure \cite{aboudi2012micromechanics}.
This introduces a multi-scale problem understood as inferring quantities of
interest on the macro-scale by means of microscopic entities
\cite{bohm_short_nodate}. To this end, computational homogenization algorithms
aim towards calculation of effective material properties by calculating
quantities of interest on the micro-scale and projecting them to higher scales
\cite{li2008introduction}. This is typically carried out using numerical
algorithms like the \textit{finite element method} (FEM) or \textit{fast Fourier
transform} (FFT) based methods. While mostly leading to more precise result
than analytical homogenization methods, a major drawback is the large
computational effort. 

A new direction in the field of engineering is opened up by machine learning
algorithms \cite{kumar2021machine}. Especially, the homogenization procedure can
be accelerated by \textit{artificial neural networks} (ANN), which are able to
approximate arbitrary Borel measurable functions \cite{hornik1989multilayer}.
Application of
ANNs to deterministic full-field homogenization can be found in
\cite{yang_deep_2018, beniwal_deep_2019, ye_deep_2019, frankel_predicting_2019}
and in the context of uncertainty quantification in
\cite{rao_three-dimensional_2020, henkes2021deep}. The concept paper
\cite{wessels2022computational}
provides a low-threshold overview of the topic. To train an ANN for
homogenization, training data is needed. In the case of microstructures, these
are usually obtained synthetically or from $\mu$CT-scans
\cite{carrara2018improved}. Although the material characteristics are more
realistic for $\mu$CT-scans than for computed microstructures
\cite{bargmann2018generation}, the available amount of image data usually is
severely  limited, as a single $\mu$CT-scan lasts several hours and hence only a
couple of them are available at reasonable cost \cite{vicente2017use}. Even
though several hundreds of microstructures can be extracted from these, this is
typically not enough for modern ANNs, which need thousands of unique
three-dimensional images during training \cite{he2021deep}. There exist
stochastic microstructure generation approaches, such as
\cite{wang2015modelling} \cite{bortolussi2018morphological},
\cite{willot2011elastic}, \cite{abdallah2016morphological},
\cite{neumann2019stochastic}, which use descriptors to model the characteristics
of the morphology. These descriptors are often non-trivial and induce a human
bias towards the properties of the microstructure.

To circumvent this problem, ANNs can be used to synthetically generate
microstructures drawn directly from the underlying generating distribution of
the $\mu$CT-scans. These special kind of ANNs are called \textit{generative
adversarial networks} (GAN). They were originally introduced for two-dimensional
images in \cite{goodfellow2014generative}, where two ANNs take part in a
two-player game of generating synthetic images and trying to discriminate them
from a set of real images, respectively. When the Nash equilibrium
\cite{kreps1989nash} in this two-player game is reached, the discriminating ANN
cannot distinguish between real and synthetic images anymore, as the quality of
the generated ones is similar to the original. The basic GAN formulation
was improved in succeeding works to overcome problems like training instability
utilizing a Wasserstein distance used in optimal transport
\cite{villani2009optimal, villani2021topics} leading to \textit{Wasserstein
GANs} \cite{arjovsky2017towards, arjovsky2017wasserstein,
gulrajani2017improved}. The current state of the art topologies are the
StyleGAN models from NVIDIA and its successors \cite{karras2019style,
karras2020analyzing, karras2021alias}. For a recent overview about standard
GAN topologies see \cite{gui2021review}. GANs are successfully engaged in
three-dimensional settings like high-energy physics \cite{carminati2018three}
and especially in medical imaging domains such as brain image reconstruction
\cite{hong20213d} and biomedical image generation \cite{wiesner2019generative}.

In the context of microstructure generation, GANs were used on 
plane porous material and rock $\mu$CT-scans using a two-dimensional StyleGAN
architecture \cite{fokina2020microstructure}. In \cite{singh2018physics}, a WGAN
was used
alongside physical constraints on the generated two-dimensional microstructures
of a synthetic dataset.  Two-dimensional microstructures of steel were produced
in \cite{lee2021virtual}.  In the three-dimensional regime, the porous media
flow properties of generated microstructures were investigated in
\cite{mosser2017reconstruction}, whereas in \cite{hsu2021microstructure} and
\cite{gayon2020pores} solid oxide fuel cell electrodes for electrochemistry
simulations where synthesized.

To the best of the authors knowledge, no attempts were made in applying the GAN
approach in the context of solid micro-mechanical homogenization. Furthermore,
state of the art GANs like StyleGAN have many parameters and take substantial
effort to train, as in the case of \cite{hsu2021microstructure}, especially in
the three-dimensional regime. The above mentioned studies often base their
network topology on ad-hoc assumptions or fail to explain their choices.
Important effects like the number of samples needed in practical applications,
where only a limited amount of micrographs are available, were not investigated.
To close these gaps, the present study therefore aims towards the following key
contributions: 

\begin{itemize} 
    \item \textbf{Three-dimensional GAN for micromechanics:} A GAN for
        three-dimensional microstructures based on $\mu$CT-scans is presented.
        The performance is evaluated by computational homogenization based on
        FFT as well as different microstructural descriptors.
    \item \textbf{Lightweight network topology:} The presented GAN utilizes
        several techniques to achieve good results tailored towards
        microstructure generation while avoiding unnecessary complexity. The GAN
        can be trained on a single workstation.
    \item \textbf{Transfer to real-world applications:} Numerical experiments
        considering important influencing factors on real-world applicability
        such as network size and number of training samples are carried out on
        two different microstructures, including a wood-plastic composite
        $\mu$CT-scan.
\end{itemize}

\noindent
The remainder of this paper is structured as follows. In
\autoref{sec:preliminaries}, a theoretical overview of the most important aspects
of this work are given. In detail, \autoref{sec:ann} introduces ANN and
\autoref{sec:gan} the main ideas of GANs. Computational homogenization
is briefly summarized in \autoref{sec:homogenization}. The proposed GAN 
is introduced and explained in \autoref{sec:mygan}. Then, two microstructures
are investigated in numerical examples. First, in \autoref{sec:example1}, a
challenging three-dimensional microstructure of spherical inclusions is
considered. Here, several
hyperparameters and techniques are discussed, such as overfitting effects and
topology choices. The resulting homogenized properties of the baseline and the
synthetic microstructures are compared.  Second, in \autoref{sec:example2}, a
$\mu$CT-scan of a wood-plastic composite is examined.  The results of this work
are discussed in \autoref{sec:end}, moreover an outlook is given. Two appendices
deal with the applicability of geometrical and physical constraints in
\ref{sec:group} and \ref{sec:PhysicGAN}, respectively, for GANs in the context
of microstructure generation and are intended for researchers who may want to
continue research along the direction of this paper.

\section{Preliminaries}
\label{sec:preliminaries} \noindent
This section introduces the basic notations of ANNs, GANs and computational
homogenization, which are essential for the developments in \autoref{sec:mygan}.

\subsection{Artificial neural networks} 
\label{sec:ann} \noindent
An ANN is a parametrized, nonlinear function composition. By the
\textit{universal function approximation theorem} \cite{hornik1989multilayer},
arbitrary Borel measurable functions can be approximated by ANN. There are
several different formulations for ANN, which can be found in standard
references such as \cite{bishop2006pattern, goodfellow2016deep,
aggarwal2018neural, geron2019hands, chollet2018deep}. Following
\cite{hauser2018principles}, most ANN formulations can be unified.
An ANN $\mathcal{N}$ is a function from an \textit{input space}
$\mathbb{R}^{d_x}$ to an \textit{output space} $\mathbb{R}^{d_y}$, defined by a
composition of nonlinear functions $\bm{h}^{(l)}$, such that 
\begin{align} 
    \mathcal{N}:
    \mathbb{R}^{d_x} &\to \mathbb{R}^{d_y} \nonumber \\ \bm{x} &\mapsto
    \mathcal{N}(\bm{x}) = \bm{h}^{(l)} \circ \ldots \circ \bm{h}^{(0)} = \bm{y},
    \quad l = 1, \ldots, n_L.  
    \label{eq:ann} 
\end{align} 
Here, $\bm{x}$ denotes an \textit{input vector} of dimension $d_x$ and $\bm{y}$
an \textit{output vector} of dimension $d_y$. The nonlinear functions
$\bm{h}^{(l)}$ are called \textit{layers} and define an $l$-fold composition,
mapping input vectors to output vectors.
Consequently, the first
layer $\bm{h}^{(0)}$ is defined as the \textit{input layer} and the last layer
$\bm{h}^{(n_L)}$ as the \textit{output layer}, such that 
\begin{equation}
    \bm{h}^{(0)} = \bm{x} \in \mathbb{R}^{d_x}, \qquad \bm{h}^{(n_L)} = \bm{y}
    \in \mathbb{R}^{d_y}. 
    \label{eq:layer} 
\end{equation} 
The layers $\bm{h}^{(l)}$ between the
input and output layer, called \textit{hidden layers}, are defined as
\begin{equation} 
    \bm{h}^{(l)} \crowr \bm{h}_{\bullet}^{(l)} = \left\{h_{\bullet,
        \eta}^{(l)}, \; \eta = 1, \ldots,
    n_{u}\right\}, \qquad h_{\bullet, \eta}^{(l)} = 
    \varphi^{(l)} \circ \;
    \phi^{(l)}\left(\bm{W}^{(l)}_{\eta} \bullet \bm{h}^{(l-1)}\right),
    \label{eq:hidden} 
\end{equation} 
where $h_{\bullet, \eta}^{(l)}$ is the $\eta$-th \textit{neural unit} of the
$l$-th layer $\bm{h}_{\bullet}^{(l)}$ and $n_u$ is the \textit{total number of
neural units per layer}, while $\bullet$ denotes a product. Following the
notation in \cite{kenningtondifferential}, the symbol $\crowr$ denotes an
abbreviation of a tuple of mathematical objects $(\mathcal{O}_1, \mathcal{O}_2,
...)$, such that $\mathcal{O} \crowr (\mathcal{O}_1, \mathcal{O}_2, ...)$.
In \autoref{eq:hidden}, the details of type-specific layers
$\bm{h}_{\bullet}^{(l)}$ are gathered in general
layers $\bm{h}^{(l)}$ from \autoref{eq:ann}. The specification follows from the
$\bullet$-operator, which denotes the operation between the \textit{weight
vector} $\bm{W}_{\eta}^{(l)}$ of the $\eta$-th neural unit in the $l$-th layer
$\bm{h}_{\bullet}^{(l)}$ and the output of the preceding layer
$\bm{h}_{\bullet}^{(l-1)}$, where the bias term is absorbed
\cite{aggarwal2018neural}. If $\bullet$ is the
ordinary matrix multiplication $\bullet = \cdot$, then the layer
$\bm{h}_{\cdot}^{(l)}$ is called \textit{dense layer}. If $\bullet$ is the
convolution or cross-correlation operation $\bullet = \star$, then the layer
$\bm{h}_{\star}^{(l)}$ is called \textit{convolutional layer}. In the context of
convolutional layers,
the weight vector $\bm{W}_{\eta}^{(l)}$ from \autoref{eq:ann} is also called
\textit{filter}, such that $n_u = n_f$ denotes the number of filters.
Furthermore, $\phi^{(l)}: \mathbb{R} \to \mathbb{R}$ is a nonlinear
\textit{activation function} and $\varphi^{(l)}$ is a function of the previous
layer, such that $\varphi^{(l)}: \bm{h}^{(l-1)} \mapsto
\varphi^{(l)}(\bm{h}^{(l-1)})$. If $\varphi^{(l)}$ is the identity function, the
layer $\bm{h}^{(l)}$ is called a \textit{feed forward layer}.  Otherwise, the
layer $\bm{h}^{(l)}$ is called \textit{residual layer}. All weight vectors
$\bm{W}_{\eta}^{(l)}$ of all layers $\bm{h}^{(l)}$ can be gathered in a single
expression, such that 
\begin{equation}
    \bm{\theta}=\left\{\bm{W}_{\eta}^{(l)}\right\}, 
    \label{eq:parameters}
\end{equation} 
where $\bm{\theta}$ inherits all parameters of the ANN $\mathcal{N}(\bm{x})$
from \autoref{eq:ann}. Consequently, the notation $\mathcal{N}(\bm{x};
\bm{\theta})$ emphasizes the dependency of the outcome of an ANN on the input on
the one hand and the current realization of the weights on the other hand. The
specific combination of layers $\bm{h}_{\bullet}^{(l)}$ from
\autoref{eq:hidden}, neural units $h_{\bullet, \eta}^{(l)}$ and activation
functions $\phi^{(l)}$ from \autoref{eq:hidden} is called \textit{topology} of
the ANN $\mathcal{N}(\bm{x}; \bm{\theta})$. The weights $\bm{\theta}$ from
\autoref{eq:parameters} are typically found by gradient-based optimization with
respect to a task-specific $\textit{loss function}$ \cite{goodfellow2016deep}.

\subsection{Generative adversarial neural networks}
\label{sec:gan} \noindent
A GAN consists of two competing ANN from \autoref{eq:ann}, namely the
\textit{generator} $\mathcal{G} \crowr \mathcal{N}(\bm{x};
\bm{\theta}) = \mathcal{N}_{\mathcal{G}}(\bm{z};
\bm{\theta}_{\mathcal{G}})$
and the \textit{discriminator} (sometimes called \textit{critic})
$\mathcal{D} \crowr \mathcal{N}(\bm{x};
\bm{\theta}) =
 \mathcal{N}_{\mathcal{D}}(\bm{I}; \bm{\theta}_{\mathcal{D}})$. The goal of the
 GAN approach is to
 create synthetic images using the generator which
have the same properties as original images from a dataset. To this end, the
generator $\mathcal{G}$ maps \textit{random input vectors} $\bm{z}$ from a
$d_x = d_Z$-dimensional input space $\mathcal{Z} \subset \mathbb{R}^{d_x} =
\mathbb{R}^{d_Z}$ called
\textit{latent space} to \textit{images} $\bm{I}$ in the $d_y = d_I$-dimensional
output \textit{synthetic image space} $\mathcal{I} \subset \mathbb{R}^{d_y} =
\mathbb{R}^{d_I}$,
such that 
\begin{align} 
    \mathcal{G}: \mathcal{Z} \subset \mathbb{R}^{d_Z} &\to \mathcal{I} \subset
    \mathbb{R}^{d_I} \nonumber \\ 
    \bm{z} &\mapsto \mathcal{G}(\bm{z}) = \bm{I}.
    \label{eq:generator} 
\end{align}
The discriminator $\mathcal{D}$ aims to distinguish between synthetic images
$\bm{I}$ and real 
images $\bm{I}_{\text{data}} \in \mathbb{D} \subset \mathcal{I}_{\text{data}}$
taken from a \textit{dataset} $\mathbb{D}$, a subset of the $d_I$-dimensional
\textit{real image space} $\mathcal{I}_{\text{data}}$, which is defined as
\begin{equation}
    \mathbb{D} = \{ \bm{I}_{\text{data}}^k, \; k = 1, ..., n_s \},
    \label{eq:dataset}
\end{equation}
where $n_s$ denotes the number of images inside the dataset called
\textit{samples}. The discriminator takes an image $\bm{I} \in \mathcal{I}$ or
$\bm{I}_{\text{data}} \in \mathcal{I}_{\text{data}}$ and gives a single number
as an output, which is a measure of similarity of the input towards the dataset
$\mathbb{D}$, such that
\begin{align} 
    \mathcal{D}: \mathcal{I} \subset \mathbb{R}^{d_I} &\to 
    \mathbb{R} \nonumber \\ 
    \bm{I} &\mapsto \mathcal{D}(\bm{I}) 
    ,
    \label{eq:discriminator} 
\end{align}
and similarly for $\bm{I}_{\text{data}} \in \mathcal{I}_{\text{data}}$.
This capability is realized by calibrating the discriminator $\mathcal{D}$ with
respect to the dataset $\mathbb{D}$ by means of a \textit{distance function}.
There are several choices for the distance between synthetic and original
images. In this work the, \textit{Wasserstein distance} $W$
is used, which is defined as
\begin{align}
    W: \mathcal{I} \times \mathcal{I}_{\text{data}} &\to \mathbb{R} \nonumber \\
    (\bm{I}, \bm{I}_{\text{data}}) &\mapsto 
    W(\bm{I}, \bm{I}_{\text{data}}) = \operatorname{inf} \mathbb{E} 
    \left[ || \bm{I} - \bm{I}_{\text{data}} ||_1 \right],
\end{align}
where $\mathbb{E}$ denotes the \textit{expectation operator}
\cite{xiu2010numerical}.
This expression is infeasible for computation. Fortunately it can be
approximated according to the \textit{Kantorovich--Rubinstein theorem}
\cite{villani2009optimal} by
\begin{equation}
    W (\bm{I}, \bm{I}_{\text{data}})\approx \tilde{W}(\bm{I},
    \bm{I}_{\text{data}}) = 
    \underset{||\mathcal{D}||_L \le1}{\operatorname{sup}} \left(
    \underset{\bm{I} \sim
    \mathcal{I}}{\mathbb{E} \left[ \mathcal{D}(\bm{I}) \right]} -
    \underset{\bm{I}_{\text{data}} \sim \mathcal{I}_{\text{data}}}{\mathbb{E}
    \left[ \mathcal{D}(\bm{I}_{\text{data}})
\right]} \right).
    \label{eq:wasserstein}
\end{equation}
Here, $||\mathcal{D}||_L \leq 1$ denotes Lipschitz-1 continuity of the
discriminator. For given
images $\bm{I}_{\text{data}} \in \mathbb{D}$ the discriminator $\mathcal{D}$ can
be trained to output the supremum in \autoref{eq:wasserstein} by maximizing the
expression with respect to a generator $\mathcal{G}$ with fixed weights
$\bm{\theta}_{\mathcal{G}}$ from \autoref{eq:parameters}. Practical
considerations such as the used \textit{loss functions} and the training
procedure are given in \autoref{sec:loss}.

\subsection{Micromechanical numerical homogenization}
\label{sec:homogenization} \noindent
The following section gives an overview of numerical homogenization in
micromechanics. A comprehensive treatment of the topic can be found in
\cite{aboudi2012micromechanics, bohm_short_nodate, li2008introduction}.

Following the notation of \cite{vondvrejc2014fft, de2017finite, zeman2017finite,
schneider2021review}, a two-phase \textit{microstructure} is represented as a
unit cell
\begin{equation}
    \mathbb{M}(x_i) = (0, \xi_1) \otimes (0, \xi_2) \otimes (0,
    \xi_3), \qquad x_i = 0,\dots, \xi_i, \qquad i = 1, 2, 3,
    \label{eq:microstructure} 
\end{equation} 
with indicator function 
\begin{align}
    \Phi(x_i) = 
    \begin{cases} 
        1, \quad x_i \in \mathbb{M}^I
        &\subseteq \mathbb{M} \\ 
        0, \quad x_i \in \mathbb{M}^M
        &\subseteq \mathbb{M}, 
    \end{cases} 
    \label{eq:microstructure2} 
\end{align}
where $\xi_i$ are dimensions and $\bm{x} = \{x_i\}$ are coordinates inside the
unit cell
$\mathbb{M} \crowr \mathbb{M}(\bm{x})$, which consists of a matrix phase
$\mathbb{M}^M$ and
inclusion phases $\mathbb{M}^I$ with $\mathbb{M} = \mathbb{M}^M \cup
\mathbb{M}^I$ and $\mathbb{M}^M \cap
\mathbb{M}^I = \emptyset$. The indicator function $\Phi$ in
\autoref{eq:microstructure2} identifies the different phases at different
coordinates $\bm{x}$ \cite{clement2013uncertainty}.

A \textit{micro-mechanical linear elastostatic problem} is defined as 
\begin{align} 
    &1. \quad
    \nabla \cdot \bm{\sigma}(\bm{x}) = \bm{0} \nonumber \\ &2. \quad
    \bm\sigma(\bm{x}) = \mathbb{C}(\bm{x}) :
    \bm\varepsilon(\bm{x}) \nonumber \\ &3. \quad
    \bm\varepsilon(\bm{x}) = \frac{1}{2}(\nabla \otimes
    \bm{u}(\bm{x}) +        (\nabla \otimes \bm{u}(\bm{x}))^T) \nonumber
    \\ &4. \quad \bar{b}(\bm{x}) \quad \text{on} \quad \partial \mathbb{M},
    \label{eq:uequilibrium} 
\end{align} 
where $\bm{\sigma}(\bm{x})$ is a second-order micro stress tensor,
$\mathbb{C}(\bm{x})$ a fourth-order micro elasticity tensor,
$\bm\varepsilon(\bm{x})$ a second-order micro strain
tensor and $\bm{u}(\bm{x})$ a first-order micro displacement tensor.
Here, \autoref{eq:uequilibrium} denotes equilibrium conditions, Hooke's law,
strain-displacement conditions and boundary conditions, respectively. For the
boundary value problem in \autoref{eq:uequilibrium} to be well posed, boundary
conditions $\bar{b}(\bm{x})$ are introduced, where $\partial \mathbb{M}$ denotes
the boundary of the corresponding unit cell $\mathbb{M}$ in
\autoref{eq:microstructure}. Possible choices are Dirichlet, Neumann or periodic
boundary conditions.

The micro elasticity tensor $\mathbb{C}(\bm{x})$ in \autoref{eq:uequilibrium}
depends on material parameters of phases $\mathbb{M}^M$ and $\mathbb{M}^I$
in \autoref{eq:microstructure2}, such that 
\begin{equation}
    \mathbb{C}(\bm{x}) = \lambda(\bm{x}) \bm{1}
    \otimes \bm{1} + 2 \mu(\bm{x})
    \mathbb{I}^{\text{sym}},
    \label{eq:elten} 
\end{equation} 
where $\lambda(\bm{x})$ is the first Lam\'{e} constant, $\mu(\bm{x})$ the
shear modulus, $\bm{1}$ the second-order unit tensor and
$\mathbb{I}^{\text{sym}}$ the symmetric fourth-order unit tensor. Solving the
boundary value problem in \autoref{eq:uequilibrium} on a microstructure
$\mathbb{M}$ in \autoref{eq:microstructure} and using an average operator on the
micro fields, one obtains corresponding effective macro fields 
\begin{equation} 
    \bar{\bm\varepsilon} = \langle \bm\varepsilon(\bm{x})
    \rangle,\quad \bar{\bm\sigma} = \langle
    \bm\sigma(\bm{x}) \rangle,\quad 
    \bar{\bm\sigma} = \bar{\mathbb{C}} :
    \bar{\bm\varepsilon},
    \label{eq:average_field} 
\end{equation}
where $\bar{\bm{\sigma}}$ denotes an effective
macro stress tensor, $\bar{\bm{\varepsilon}}$ an 
effective macro strain tensor and $\bar{\mathbb{C}}$ an
effective macro elasticity tensor. The average operator $\langle \bullet
\rangle$ in \autoref{eq:average_field} is defined as \begin{equation}
    \langle \bullet \rangle = \displaystyle\frac{1}{\mathbb{M}}
\displaystyle\int_{\mathbb{M}} (\bullet) \mathrm{d}\mathbb{M}.  \end{equation}
The macro and micro stress and strain fields need to satisfy the
\textit{Hill-Mandel condition}: 
\begin{equation} \langle
    \bm{\sigma}(\bm{x}) : \bm{\varepsilon}(\bm{x}) \rangle =
    \langle \bm{\sigma}(\bm{x}) \rangle : \langle
    \bm{\varepsilon}(\bm{x}) \rangle.  
    \label{eq:Hill_Mandel}
\end{equation}
In numerical or full-field homogenization, the microstructure $\mathbb{M}$ from
\autoref{eq:microstructure} is given explicitly and the boundary problem in
\autoref{eq:uequilibrium} is solved by usage of e.g., FEM or FFT based methods,
usually utilizing periodic boundary conditions. To obtain all components of the
effective elasticity tensor in \autoref{eq:average_field}, six uniaxial macro
strains $\bar{\varepsilon}$ have to be applied on the microstructure
$\mathbb{M}$, such that 
\begin{equation}
    \bar{\underline{\underline{\mathbb{C}}}} =
    [\bm{\bar{\underline{{\sigma}}}}_1, 
        \bm{\bar{\underline{{\sigma}}}}_2,
        \bm{\bar{\underline{{\sigma}}}}_3, 
        \bm{\bar{\underline{{\sigma}}}}_4,
    \bm{\bar{\underline{{\sigma}}}}_5, 
\bm{\bar{\underline{{\sigma}}}}_6], \quad
\bm{\bar{\underline{{\sigma}}}}_i = 
\langle \underline{\underline{\mathbb{C}}} \;
    \underline{\bar{\bm{\varepsilon}}}_i \rangle, \; i = 1,\cdots,6,
    \label{eq:full_homo}
\end{equation}
\begin{equation*}
\begin{aligned}
    \bm{\bar{ \underline\varepsilon}}_1 = \left[1, 0, 0, 0, 0,
    0\right]^T, \quad \bm{\bar{ \underline\varepsilon}}_2 = \left[0, 1, 0,
    0, 0, 0\right]^T, \quad \bm{\bar{ \underline\varepsilon}}_3 = \left[0,
    0, 1, 0, 0, 0\right]^T,\\ \bm{\bar{ \underline\varepsilon}}_4 =
    \left[0, 0, 0, 1, 0, 0\right]^T, \quad \bm{\bar{
    \underline\varepsilon}}_5 = \left[0, 0, 0, 0, 1, 0\right]^T, \quad
    \bm{\bar{ \underline\varepsilon}}_6 = \left[0, 0, 0, 0, 0,
    1\right]^T,\\
 \end{aligned}
\end{equation*}
where $\underline{\bullet}$ denotes the vector representation of a second-order
tensor and $\underline{\underline{\bullet}}$ the matrix representation of a
fourth-order tensor. For details the reader is referred to
\cite{li2008introduction}.

The microstructure $\mathbb{M}$ from \autoref{eq:microstructure} can be
characterized by \textit{n-point correlation functions} \cite{singh2018physics}.
In this work the first two n-point correlation functions are used to describe
the properties of $\mathbb{M}$. In contrast to stochastic microstructure
generation approaches, where such descriptors are used explicitly during
synthesis, in this work these functions are only used to evaluate
some characteristics of the microstructure during post-processing. 1-point
correlation function is defined as
\begin{align}
    p_1 = \underset{\bm{x} \sim \mathbb{M}} {\mathbb{E}} 
    \left[ \Phi(\bm{x}) \right], 
    \label{eq:1point}
\end{align}
with indicator function $\Phi(\bm{x})$ from \autoref{eq:microstructure2},
which is equivalent to the inclusion volume fraction of the microstructure. The
2-point correlation function is defined as
\begin{align}
    p_2 = \underset{\bm{x}_1, \bm{x}_2 \sim \mathbb{M}}{\mathbb{E}} 
    \left[ \Phi(\bm{x}_1) \Phi(\bm{x}_2) \right],
    \label{eq:2point}
\end{align}
which includes information about the shape and distribution of inclusions.

The numerical homogenization problem in \autoref{eq:full_homo} becomes
computational demanding, if the dimensions $\xi_i$ of the microstructure
$\mathbb{M}$ from \autoref{eq:microstructure} get high. Consequently, an
efficient surrogate model using ANN was proposed in \cite{henkes2021deep}, which
is able to solve the forward problem magnitudes faster. One drawback in the
above mentioned work is the number of training $\mu$CT-scans needed, which in
practice are often not available due to financial constraints. In the following,
this gap is closed by proposing a fast generative algorithm based on ANN, which
does not include physical descriptors of any kind and is applicable to arbitrary
materials. 

\section{GAN for three-dimensional microstructure generation}
\label{sec:mygan} \noindent
In the following, the network topologies of the ANNs as well as the loss
functions used in this work are presented. Furthermore, evaluation metrics are
discussed. 

\subsection{Generator topology} 
\label{sec:topology_generator} \noindent
The generator $\mathcal{G}$ from \autoref{eq:generator} maps a random input
vector $\bm{z} \in \mathcal{Z}$ to an image $\bm{I} \in \mathcal{I}$. In this
work, an independent multivariate normal random input vector is used, where
every component of the $d_Z$-dimensional vector is normally distributed with
mean $\mu$ and standard deviation $\sigma$, such that 
\begin{equation}
    \bm{z} =  (z_m \sim
    \mathcal{N}(\mu, \sigma^2)), \quad m = 1, ..., d_Z,
    \label{eq:random_input}
\end{equation}
where in contrast to the previous usage of $\mathcal{N}$ as an ANN, here, by
convention, $\sim \mathcal{N}(\mu, \sigma^2)$ denotes \textit{normally
distributed} \cite{xiu2010numerical}. In the following, the meaning will be
clear from the context. Similar to the StyleGAN architecture
\cite{karras2019style}, the latent vector $\bm{z}$ is first mapped to a
$d_W$-dimensional nonlinear transformed distribution $\bm{w} \in \mathcal{W}$.
This is carried out by a \textit{mapping network} $\mathcal{N}_{\text{map}}$,
such that
\begin{align} 
    \mathcal{N}_{\text{map}}:
    \mathcal{Z} \subset \mathbb{R}^{d_Z} &\to \mathcal{W} \subset
    \mathbb{R}^{d_W} \nonumber \\
    \bm{z} &\mapsto
    \mathcal{N}_{\text{map}}(\bm{z}) = \bm{h}_{\cdot}^{(l)} \circ \ldots \circ
    \bm{h}_{\cdot}^{(0)} = \bm{w},
    \quad l = 1, \ldots, n_{Lw}.
    \label{eq:ann_map} 
\end{align} 
The mapping network consists of $n_{L}^{w}$ dense layers $\bm{h}_{\cdot}^{(l)}$
from \autoref{eq:hidden} with \textit{leaky ReLU} activation function
\cite{maas2013rectifier}
\begin{align}
    \phi_{\text{LReLU}}(x) = 
    \begin{cases} 
        x, \quad &x \geq 0 \\
        \gamma x, \quad &x < 0, \quad \gamma \in \mathbb{R}_{+},
    \end{cases} 
    \label{eq:leakyrelu} 
\end{align}
for some input $x \in \mathbb{R}$ and a constant $\gamma \in \mathbb{R}_+$.
Then, the nonlinear distribution $\bm{w}$ is dimensionally upsampled by
\textit{transpose residual convolution blocks} $\mathcal{N}_{\text{ResConvT}}$
consisting of strided transpose convolutional layers
\cite{zeiler2010deconvolutional} $\bm{h}_{\star T}^{(l)}(\bm{x})$ with kernel
size $n_k \times n_k \times n_k$ and strided transpose convolutional layers
$\bm{h}_{\star T}^{(l)}(\bm{w})$ with kernel size $1 \times 1 \times 1$. Here,
the $1 \times 1 \times 1$ convolutional layer $\bm{h}_{\star}^{(l)}(\bm{w})$
acts as an additive residual layer, adding the upsampled nonlinear random vector
$\bm{w}$ to every $\mathcal{N}_{\text{ResConvT}}$ block, such that
\begin{equation}
    \mathcal{G} \left( \bm{z} \right)
    = \bm{h}_{\star}^{\operatorname{tanh}} \circ
    \mathcal{N}_{\text{ResConvT}}^{(l)} \circ \ldots \circ
        \mathcal{N}_{\text{ResConvT}}^{(1)} \circ
        \mathcal{N}_{\text{map}} 
        \left( \bm{z} \right) = \bm{I}, \quad l = 1, ..., n_L^{\mathcal{G}},
\end{equation}
with
\begin{equation}
    \mathcal{N}_{\text{ResConvT}}^{(l)} \left( \bm{x} \right) = 
    \phi_{\text{LReLU}}
    \left(
        \bm{h}_{\star T}^{(l)} (\bm{w}) + \bm{h}_{\star T}^{(l)} \left( \bm{x}
        \right)
    \right),
    \label{eq:block_g}
\end{equation}
where $n_L^{\mathcal{G}}$ denotes the number of blocks of the generator.
The strides for the $n_k \times n_k \times n_k$ layers are chosen as $2$ to
double the dimension of the layer output from block to block.
The strides for the $1 \times 1 \times 1$ residual layers
$\bm{h}_{\star}^{(l)}(\bm{w})$ have to be chosen in such a way that they match
the dimensions of the $\bm{h}_{\star T}^{(l)}(\bm{x})$ layers. As the dimensions
grow in every block, successively larger strides have to been chosen for
upsampling $\bm{w}$. The number of
blocks defines the final output dimension $d_I$ from \autoref{eq:generator} and
therefore the dimension of the output image $\bm{I}$. All convolutional blocks
use leaky ReLU activation functions from \autoref{eq:leakyrelu}, except the last
one.  The last layer $\bm{h}_{\star}^{\operatorname{tanh}}$ uses a modified
$\operatorname{tanh}$ activation function, which constraints the output values
to lie between the minimum and maximum voxel values, $\lambda_{min}$ and
$\lambda_{max}$ respectively, apparent in the dataset $\mathbb{D}$ from
\autoref{eq:dataset}. It is defined as
\begin{equation} 
\bm{h}_{\star}^{\operatorname{tanh}}(\bm{x}) = 
\operatorname{tanh}(\bm{x} + \bm{1}) \frac{\lambda_{max} -
\lambda_{min}}{2} + \lambda_{min}\bm{1},
    \label{eq:generator_activation} 
\end{equation} 
with unit matrix $\bm{1}$ and
\begin{align}
    \lambda_{min} &= \operatorname{min} v \in \bm{I}_{\text{data}} \; \forall \;
    \bm{I}_{\text{data}} \in \mathbb{D}, \\ \nonumber 
    \lambda_{max} &= \operatorname{max} v \in \bm{I}_{\text{data}} \; \forall \;
    \bm{I}_{\text{data}} \in \mathbb{D}, 
\end{align}
where $v$ denotes the value of a single voxel in the image
$\bm{I}_{\text{data}}$. As a slight abuse of notation, the layer
$\bm{h}_{\star}^{\operatorname{tanh}}(\bm{x})$ includes its activation function,
as opposed to the general definition given in \autoref{eq:hidden}, to shorten
notation. An illustration of the generator topology is shown in
\autoref{fig:generator}.

\subsection{Discriminator topology} 
\label{sec:topology_discriminator} 
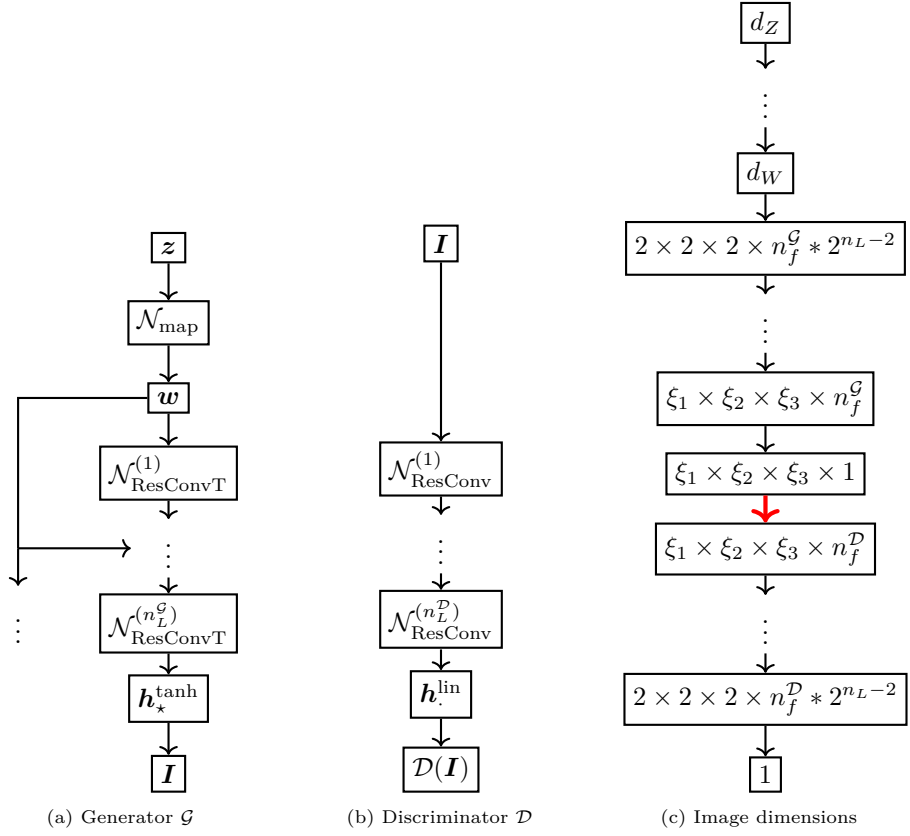
\begin{figure}[htb]
    \centering
    \begin{subfigure}[b]{0.3\textwidth}
        \centering
        \begin{tikzpicture}[]
            \node (z) at (0.0, 0.0) [rectangle, draw, thick] 
            {$\bm{z}$};

            \node (map) at (0.0, -1.0) [rectangle, draw, thick] 
            {$\mathcal{N}_{\text{map}}$};

            \draw[->, thick] (z) to (map);

            \node (w) at (0.0, -2.0) [rectangle, draw, thick] 
            {$\bm{w}$};

            \draw[->, thick] (map) to (w);

            \node (resconvt_1) at (0.0, -3.0) [rectangle, draw, thick] 
            {$\mathcal{N}_{\text{ResConvT}}^{(1)}$};

            \draw[->, thick] (w) to (resconvt_1);

            \draw[thick] (w) -- (-2.0, -2.0) -- (-2.0, -4.0);

            \draw[->, thick] (-2.0, -4.0) -- (-0.5, -4.0);

            \draw[->, thick] (-2.0, -4.0) -- (-2.0, -4.5);

            \node[draw=none] (dots) at (0.0, -4.0) {$\vdots$};

            \node (dots_2) at (-2.0, -5.0) [] {$\vdots$};

            \draw[->, thick] (resconvt_1) to (dots);

            \node (resconvt_2) at (0.0, -5.0) [rectangle, draw, thick] 
            {$\mathcal{N}_{\text{ResConvT}}^{(n_L^{\mathcal{G}})}$};

            \draw[->, thick] (dots) to (resconvt_2);

            \node (tanh) at (0.0, -6.0) [rectangle, draw, thick] 
            {$\bm{h}_{\star}^{\operatorname{tanh}}$};

            \draw[->, thick] (resconvt_2) to (tanh);

            \node (I) at (0.0, -7.0) [rectangle, draw, thick] 
            {$\bm{I}$};

            \draw[->, thick] (tanh) to (I);
        \end{tikzpicture}
        \caption{Generator $\mathcal{G}$}
        \label{fig:generator}
    \end{subfigure}
    \hfill
    \begin{subfigure}[b]{0.3\textwidth}
        \centering
        \begin{tikzpicture}[]
            \node (I) at (0.0, 0.0) [rectangle, draw, thick] 
            {$\bm{I}$};

            \node (resconv_1) at (0.0, -3.0) [rectangle, draw, thick] 
            {$\mathcal{N}_{\text{ResConv}}^{(1)}$};

            \draw[->, thick] (I) to (resconv_1);

            \node[draw=none] (dots) at (0.0, -4.0) {$\vdots$};

            \draw[->, thick] (resconv_1) to (dots);

            \node (resconv_2) at (0.0, -5.0) [rectangle, draw, thick] 
            {$\mathcal{N}_{\text{ResConv}}^{(n_L^{\mathcal{D}})}$};

            \draw[->, thick] (dots) to (resconv_2);

            \node (lin) at (0.0, -6.0) [rectangle, draw, thick] 
            {$\bm{h}_{\cdot}^{\operatorname{lin}}$};

            \draw[->, thick] (resconvt_2) to (lin);

            \node (DI) at (0.0, -7.0) [rectangle, draw, thick] 
            {$\mathcal{D}(\bm{I})$};

            \draw[->, thick] (lin) to (DI);
        \end{tikzpicture}
        \caption{Discriminator $\mathcal{D}$}
        \label{fig:discriminator}
    \end{subfigure}
    \hfill
    \begin{subfigure}[b]{0.3\textwidth}
        \centering
        \begin{tikzpicture}[]
            \node (z) at (0.0, 0.0) [rectangle, draw, thick] 
            {$d_Z$};

            \node[draw=none] (dots) at (0.0, -1.0) {$\vdots$};

            \draw[->, thick] (z) to (dots);

            \node (w) at (0.0, -2.0) [rectangle, draw, thick] 
            {$d_W$};

            \draw[->, thick] (dots) to (w);

            \node (g1) at (0.0, -3.0) [rectangle, draw, thick] 
            {$2 \times 2 \times 2 \times n_f^{\mathcal{G}} * 2^{n_L - 2}$};

            \draw[->, thick] (w) to (g1);

            \node[draw=none] (dots2) at (0.0, -4.0) {$\vdots$};

            \draw[->, thick] (g1) to (dots2);

            \node (g2) at (0.0, -5.0) [rectangle, draw, thick] 
            {$\xi_1 \times \xi_2 \times \xi_3 \times n_f^{\mathcal{G}}$};

            \draw[->, thick] (dots2) to (g2);

            \node (image) at (0.0, -6.0) [rectangle, draw, thick] 
            {$\xi_1 \times \xi_2 \times \xi_3 \times 1$};

            \draw[->, thick] (g2) to (image);

            \node (d1) at (0.0, -7.0) [rectangle, draw, thick] 
            {$\xi_1 \times \xi_2 \times \xi_3 \times n_f^{\mathcal{D}}$};

            \draw[->, ultra thick, red] (image) to (d1);

            \node[draw=none] (dots3) at (0.0, -8.0) {$\vdots$};

            \draw[->, thick] (d1) to (dots3);

            \node (d2) at (0.0, -9.0) [rectangle, draw, thick] 
            {$2 \times 2 \times 2 \times n_f^{\mathcal{D}} * 2^{n_L - 2}$};

            \draw[->, thick] (dots3) to (d2);

            \node (DI) at (0.0, -10.0) [rectangle, draw, thick] 
            {$1$};

            \draw[->, thick] (d2) to (DI);
        \end{tikzpicture}
        \caption{Image dimensions}
        \label{fig:image}
    \end{subfigure}
    \caption{The network topology of the generator
        $\mathcal{G}$ from \autoref{eq:generator} and the discriminator
    $\mathcal{D}$ from \autoref{eq:discriminator}. (a) For the generator, the
spatial image dimensions go up from $\mathcal{N}_{\text{ResConvT}}^{(1)}$ to
$\mathcal{N}_{\text{ResConvT}}^{(n_L^{\mathcal{G}})}$, whereas the number of
filters $n_f^{\mathcal{G}}$ goes down. (b) The opposite is true for the
discriminator. Here, the
spatial image dimensions go down from $\mathcal{N}_{\text{ResConv}}^{(1)}$ to
$\mathcal{N}_{\text{ResConv}}^{(n_L^{\mathcal{D}})}$, whereas the number of
filters $n_f^{\mathcal{G}}$ goes up. (c) The relation between spatial dimension
of the generated image and filter size is illustrated. The generated image has
dimensions $\xi_1 \times \xi_2 \times \xi_3 \times 1$, where $\xi_i$ denotes the
edge length of the microstructure from \autoref{eq:microstructure}. In this
work, an arbitrary factor $2^{n_L - 2}$ has been chosen for the maximum number
of filters in both networks, which showed good performance. The power of two was
chosen to accelerate the calculations on modern TensorCore GPUs using mixed
precision. The transition between generator and discriminator is depicted by
the red arrow.}
    \label{fig:topology}
\end{figure}

\noindent
The discriminator $\mathcal{D}$ from \autoref{eq:discriminator} maps an image
$\bm{I}$ to a real number $\mathcal{D}(\bm{I})$, which is then incorporated into
the Wasserstein distance $\mathcal{L}_W$ in \autoref{eq:wasserstein}. The image
$\bm{I}$ is downsampled by means of their spatial dimension utilizing
\textit{residual convolution blocks} $\mathcal{N}_{\text{ResConv}}$ consisting
of convolutional layers \cite{zeiler2010deconvolutional}
$\bm{h}_{\star}^{(l)}(\bm{x})$ with kernel size $3 \times 3 \times 3$ and $1
\times 1 \times 1$,
followed by \textit{average pooling} ($\operatorname{AvgPool}$)
\cite{goodfellow2016deep} for dimension
reduction. Here, the $1 \times 1 \times 1$ convolutional layer
$\bm{h}_{\star}^{(l)}(\bm{w})$ acts as an additive residual layer, providing the
output of the preceding layer $\bm{h}_{\star}^{(l - 1)}(\bm{x})$ in every block,
such that
\begin{equation}
    \mathcal{D} \left( \bm{I} \right)
    = \bm{h}_{\cdot}^{\operatorname{lin}} \circ
    \mathcal{N}_{\text{ResConv}}^{(l)} \circ \ldots
    \circ \mathcal{N}_{\text{ResConv}}^{(1)} \left(\bm{I}
    \right), \quad l = 1, ..., n_L^{\mathcal{D}},
\end{equation}
with
\begin{equation}
    \mathcal{N}_{\text{ResConv}}^{(l)} \left( \bm{x} \right) = 
    \phi_{\text{LReLU}} \circ 
    \operatorname{AvgPool} 
    \circ \;
    \bm{h}_{\star}^{(l)}
    \circ 
    \left( \bm{h}_{\star}^{(l - 1)} + \bm{h}_{\star}^{(l)} \left( \bm{x} \right)
    \right),
    \label{eq:block_d}
\end{equation}
where $n_L^{\mathcal{D}}$ denotes the number of blocks for the discriminator.
All convolutional blocks use leaky ReLU activation functions
\autoref{eq:leakyrelu}. The last layer $\bm{h}_{\cdot}^{\operatorname{lin}}$ is
a dense layer and uses a linear activation function. An illustration of the
discriminator topology is shown in \autoref{fig:discriminator}. 

\FloatBarrier
\subsection{Loss function and training} 
\label{sec:loss} \noindent
The generator $\mathcal{G}$ from \autoref{eq:generator} and the discriminator
$\mathcal{D}$ from \autoref{eq:discriminator} have different loss functions,
which are minimized during optimization of their respective weights
$\bm{\theta}_{\mathcal{G}}$ and $\bm{\theta}_{\mathcal{D}}$ from
\autoref{eq:parameters}. The overall goal for the
generator is to produce synthetic images $\bm{I}$ from \autoref{eq:generator},
which have a small Wasserstein distance $W$ from \autoref{eq:wasserstein} with
respect to a dataset $\mathbb{D}$ from \autoref{eq:dataset}. 

\subsection{Discriminator loss} \noindent
The discriminator loss $\mathcal{L}_{\mathcal{D}}$ consists of two terms.
The first, $\mathcal{L}_{\tilde{W}}$, incorporates the approximated
Wasserstein distance ${\tilde{W}}$ from \autoref{eq:wasserstein},
the second, $\mathcal{L}_{\text{GP}}$, is a gradient penalty term. The
discriminator loss $\mathcal{L}_{\mathcal{D}}$ is therefore
\begin{equation}
    \mathcal{L}_{\mathcal{D}} = \mathcal{L}_{\tilde{W}} +
    \mathcal{L}_{\text{GP}},
    \label{eq:loss_d}
\end{equation}
with
\begin{equation}
\mathcal{L}_{\tilde{W}} = 
    \underset{\bm{I} \sim
    \mathcal{I}}{\mathbb{E}} \left[ \mathcal{D}(\bm{I}) \right] -
    \underset{\bm{I}_{\text{data}} \sim \mathcal{I}_{\text{data}}}{\mathbb{E}}
    \left[ \mathcal{D}(\bm{I}_{\text{data}})
    \right].
    \label{eq:loss_d_w}
\end{equation}
The gradient penalty term $\mathcal{L}_{GP}$ ensures Lipschitz-1 continuity,
which is defined as
\begin{equation} \mathcal{L}_{GP} = 
    \lambda \underset{\hat{\bm{I}} \sim 
    \hat{\mathcal{I}}}{\mathbb{E}}
    \left[ \left( || \nabla_{\hat{\bm{x}}}
    \mathcal{D}(\hat{\bm{I}}) ||_2 - 1 \right)^2 \right].
\end{equation}
where $\hat{\mathcal{I}}$ is a uniform distribution interpolated
between $\mathcal{I}_{\text{data}}$ from \autoref{eq:dataset} and $\mathcal{I}$
from \autoref{eq:generator} and $\lambda$ is a weight term. The Lipschitz-1
constraint is necessary due to the Kantorovich--Rubinstein theorem.  Further
details can be found in \cite{gulrajani2017improved}. The minimization of the
loss function increases the Wasserstein distance between synthetic images
$\bm{I}$ and real images $\bm{I}_{\text{data}}$. The real images are
standardized before being provided to the discriminator, such that 
$\bm{I}_{\text{data}} \sim \bm{\mathcal{N}}(\bm{0}, \bm{1})$ is a multivariate
standard normal distribution \cite{xiu2010numerical}.  

\subsection{Generator loss} \noindent
The generator loss $\mathcal{L}_{\mathcal{G}}$ is defined as
\begin{equation}
    \mathcal{L}_{\mathcal{G}} = - \underset{\bm{I} \sim
    \mathcal{I}}{\mathbb{E}} \left[ \mathcal{D}(\bm{I}) \right].
    \label{eq:loss_g}
\end{equation}
This is the negative version of the first term in \autoref{eq:loss_d_w}. By
minimizing towards the opposite direction, the generator successively minimizes
the Wasserstein distance for its synthetic images $\bm{I}$.  During training,
the generator adapts the statistics of the synthetic images $\bm{I}$ towards the
statistics of the real images, which are follows a standard normal
distribution $\bm{I}_{\text{data}} \sim \bm{\mathcal{N}}(\bm{0}, \bm{1})$. These
statistics can be monitored during training and may aid as a convergence
criterion.

\subsection{Training} \noindent
The overall goal is to train the
generator $\mathcal{G}$ to minimize the Wasserstein distance $W$ for its
generated images $\bm{I}$. To this end, the generator $\mathcal{G}$
produces synthetic images $\bm{I}$ from random input $\bm{z}$, while the
discriminator $\mathcal{D}$ measures the distance between the generated
and real images by means of the Wasserstein metric. The generator is then
updated in the distance minimizing gradient direction, whereas the discriminator
is updated in the distance maximizing gradient direction, such that
\begin{equation}
    \argmin_{\bm{\theta}_{\mathcal{G}}} \mathcal{L}_{\mathcal{G}}, 
    \qquad
    \argmin_{\bm{\theta}_{\mathcal{D}}} \mathcal{L}_{\mathcal{D}},
        \label{eq:opt} 
\end{equation}
which is typically carried out by stochastic gradient descent using a
\textit{learning rate} $\alpha$ and a batch size $n_B$.  The whole training
process can only be successfully carried out if the discriminator is trained to
give successively better approximations of the true Wasserstein metric, while
the generator is trained to produce images which minimize the Wasserstein
metric. Because the discrimination capability of the discriminator depends on
the generation capability of the generator, the training process is highly
dynamic. Care has to be taken, such that the approximated Wasserstein distance
in \autoref{eq:wasserstein} is always close enough on the true Wasserstein
distance to give the generator meaningful feedback. This can be achieved by
choosing a higher learning rate during optimization for the discriminator
compared to the generator \cite{gulrajani2017improved}. The training algorithm
is summarized in \autoref{algo:training}.

\algsetup{indent=2em} 
\begin{algorithm}[htb] 
    \renewcommand{\thealgorithm}{1}
    \caption{GAN training} 
    \label{algo:training} 
    \begin{algorithmic}[0]
        \REQUIRE Dataset $\mathbb{D}$ from \autoref{eq:dataset}
        \REQUIRE Generator $\mathcal{G}$ from \autoref{eq:generator}
        \REQUIRE Discriminator $\mathcal{D}$ from \autoref{eq:discriminator}
        \FOR{i in $n_{\text{iter}}$}
        \STATE Sample random input vector $\bm{z}$ from
        \autoref{eq:random_input}
        \STATE Generate image $\mathcal{G}(\bm{z}) = \bm{I}$
        \STATE Sample real image $\bm{I}_{\text{data}}$ from dataset
        $\mathbb{D}$
        \STATE Calculate generator loss $\mathcal{L}_{\mathcal{G}}$ from
        \autoref{eq:loss_g}
        \STATE Calculate discriminator loss $\mathcal{L}_{\mathcal{D}}$ from
        \autoref{eq:loss_d}
        \STATE Update weights $\bm{\theta}_{\mathcal{G}}$ and
        $\bm{\theta}_{\mathcal{D}}$ from \autoref{eq:parameters} as in
        \autoref{eq:opt} 
        \ENDFOR
    \end{algorithmic} 
\end{algorithm}

\subsection{Evaluation metric} 
\label{sec:metrics}\noindent
To quantify the quality of the generated synthetic images $\mathcal{G}(\bm{z}) =
\bm{\mathcal{I}}$ from \autoref{eq:generator}, three criterion metrics are
considered, namely the
\begin{itemize}
    \item scalar-valued 1-point correlation function $p_1$ from
        \autoref{eq:1point} 
    \item vector-valued 2-point correlation function $p_2$ from
        \autoref{eq:2point}
    \item scalar-valued effective elasticity $\bar{\mathbb{C}}_{11}$-tensor
        component of the tensor $\bar{\mathbb{C}}$ from
        \autoref{eq:average_field}
\end{itemize}
The 1-point correlation function $p_1$ measures the volume fraction of the
inclusion phase $\mathbb{M}^I$, whereas the 2-point correlation function $p_2$
is characteristic for the distribution and shapes of the inclusions. Finally,
the effective elasticity $\bar{\mathbb{C}}_{11}$-tensor component of the tensor
$\bar{\mathbb{C}}$ measures the homogenized mechanical properties of the
microstructure in one loading direction. 
For these three evaluation criteria, a scalar-valued error function can be
defined, which measures the relative error with respect to a baseline. The
baseline is calculated from a training dataset $\mathbb{D}$.

In this work, the mean relative error $\mathcal{E}$ is used, which is
defined as
\begin{equation}
    \mathcal{E}_{\bullet}(\bullet) = 
    \frac{1}{n_s}
    \displaystyle\sum_{i=1}^{n_s}
    \frac{||\hat{{\bullet}}_i - {\bullet}_i||_2}{||\hat{{\bullet}}_i||_2},
    \label{eq:error}
\end{equation}
for some input $\bullet$ and baseline $\hat{\bullet}$.
Here, $\bullet$ equals $\bullet = p_1$ for the 1-point correlation function,
$\bullet = p_2$ for the 2-point correlation function or $\bullet =
\bar{\mathbb{C}}_{11}$ for the effective elasticity
$\bar{\mathbb{C}}_{11}$-tensor component.

\FloatBarrier
\section{Example 1: Spherical inclusions} 
\label{sec:example1}
\begin{figure}[htb]
    \centering
    \begin{subfigure}[b]{0.45\textwidth}
        \centering
        \includegraphics[trim=10cm 4cm 10cm 8cm,clip,width=1.0\textwidth]
        {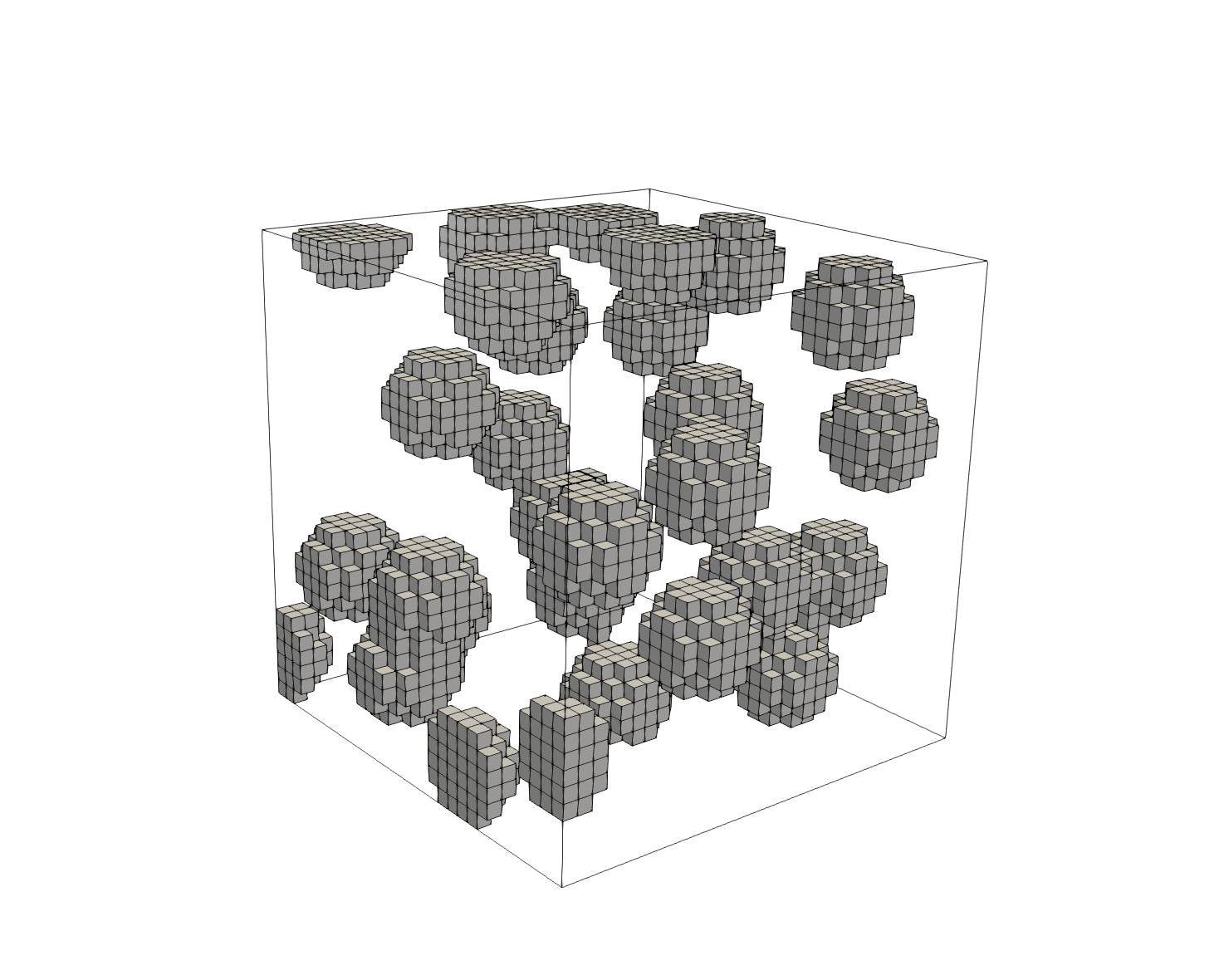}
        \caption{3D representation}
        \label{fig:micro_3d}
    \end{subfigure}
    \hfill
    \begin{subfigure}[b]{0.45\textwidth}
        \centering
        \includegraphics[width=1.0\textwidth]{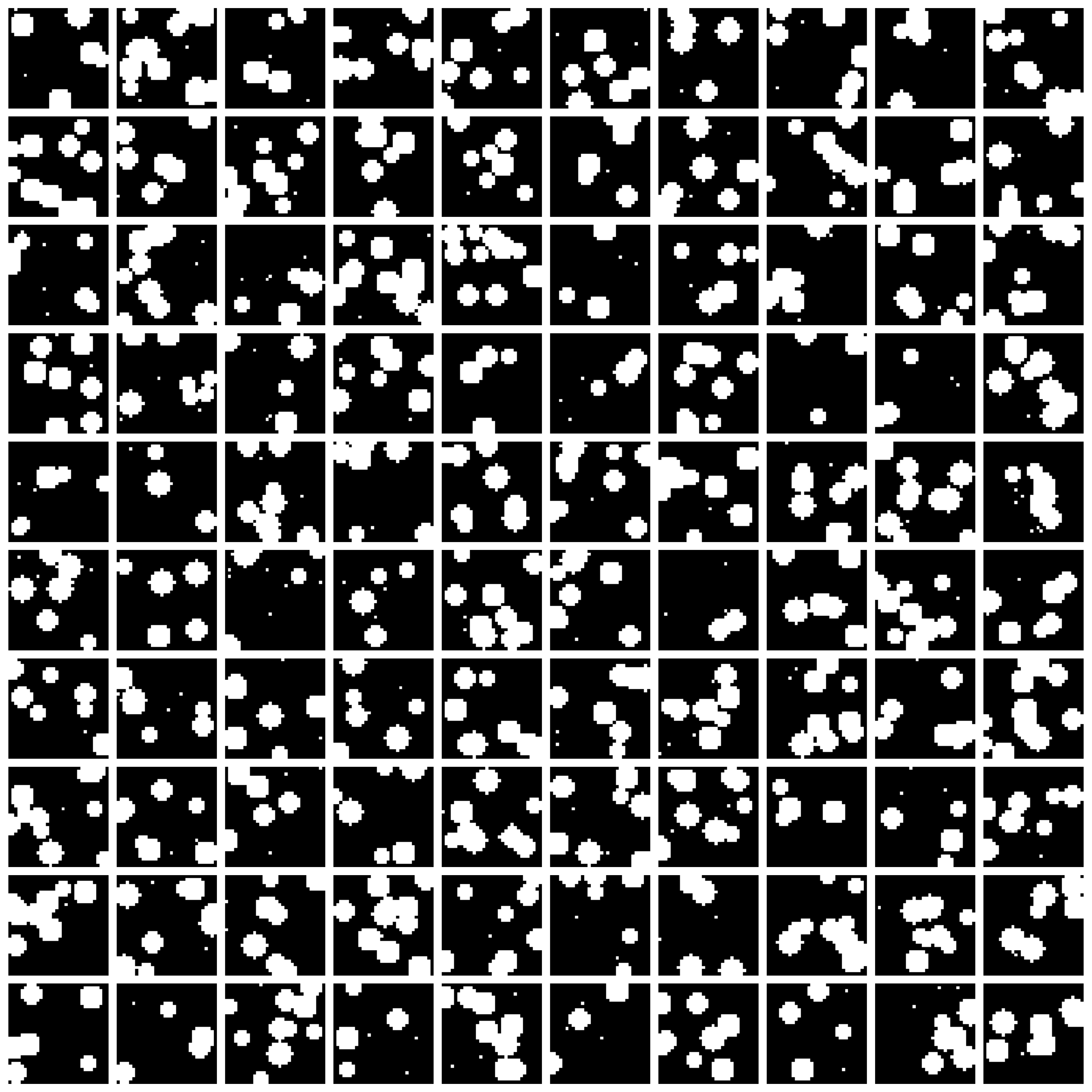}
        \caption{2D representation}
        \label{fig:micro_2d}
    \end{subfigure}
    \caption{\textbf{Example 1: Training microstructure.} Spherical inclusions
        embedded in a matrix material as used throughout \autoref{sec:example1}.
    (a) Three-dimensional representation of a single microstructure with
suppressed matrix. (b) Center plane slices of $1 \times 10^2$ three-dimensional
microstructures.}
    \label{fig:microstructure_real_100}
\end{figure}

\noindent
The first numerical example deals with microstructures consisting of spherical
inclusions embedded into a matrix material, where the linear elastic material
parameters of both constituents differ from each other. The single
microstructures are represented in a voxel format with $32^3$ voxels, where the
radius of the spheres is $4$ voxels. The inclusion volume fraction from
\autoref{eq:1point} is $p_1 = 0.2$.  The Young's moduli for the matrix and
inclusion are $E = [10^3, 10^4]$, respectively. The Poisson's ratios are set as
$\nu = [0.4, 0.1]$ in the same manner. These values can be translated to the
material parameters $\lambda$ and $\mu$ given in \autoref{eq:elten} in the well
known way, see e.g., \cite{henkes2021physics}.  The microstructure is
illustrated in \autoref{fig:microstructure_real_100}. 

For all experiments, the same hyperparameters are used, if not indicated
otherwise. The experiments were carried out on a single workstation utilizing a
GeForce RTX 3090 GPU and TensorFlow 2 \cite{tensorflow2015-whitepaper}. To
take advantage of the TensorCore \cite{markidis2018nvidia} architecture of the
GPU, mixed precision is used.  Here, only crucial numerical values, such as
weights, biases and loss functions, are saved in single precision, whereas most
other variables, such as inputs,  are converted to half precision. Both networks
are optimized using the stochastic gradient Nadam optimizer, a variant of the
Adam optimizer utilizing Nesterov momentum \cite{dozat2016incorporating}.  The
hyperparameters of this optimizer are the learning rate $\alpha$ and two
\textit{moment} terms $\beta_1, \beta_2$, which are set to $\beta_1 = 0.9,
\beta_2 = 0.999$ for all networks. For additional numerical stability, global
norm gradient clipping is used \cite{aggarwal2018neural}. The learning rate is
limited by the usage of mixed precision due to numerical under- and overflow. In
this context, the learning rate is chosen as large as possible. For all
experiments, the learning
rate for the discriminator is $\alpha_{\mathcal{D}} = 1 \times 10^{-4}$ and for
the generator  $\alpha_{\mathcal{G}} = 5 \times 10^{-5}$. This is because the
discriminator has to be trained more than the generator, as explained in
\autoref{sec:loss}.  Furthermore, the gradient penalty factor $\lambda$ from
\autoref{eq:loss_d} is
set to $\lambda = 10$.  To avoid instabilities, as reported in
\cite{karras2021alias}, the mapping network $\mathcal{N}_{\text{map}}$ from
\autoref{eq:ann_map} of the generator uses $n_L^w = 8$ layers and is updated
with a reduced learning rate, such that $\alpha_{\mathcal{N}_{\text{map}}} = 0.1
\times \alpha_{\mathcal{G}}$. The mean and standard deviation of the normal
input random vector components to the generator are chosen as $\mu = 0$ and
$\sigma = 1$. The \textit{batch size} $n_B$ for both the generator and
discriminator was chosen to be $n_B = 8$ to match the relatively small learning
rate and use the advantages of TensorCores. All networks are initialized
orthogonally \cite{saxe2013exact}.  The homogenization as outlined in
\autoref{sec:homogenization} is carried out by FFT based on the framework
presented in \cite{vondvrejc2014fft, de2017finite, zeman2017finite}.  The
2-point correlation function from \autoref{eq:2point} is calculated using FFT
and a radial norm using a custom TensorFlow implementation of
\cite{Gostick2019}.

The goal of this section is to show the influence of specific hyperparameters on
the ability of the proposed GAN approach to create synthetic microstructures.
The investigated hyperparameters are crucial for the performance of the proposed
GAN, both by means of precision and training time. A summary of the most
important hyperparameters used in this work can be found in
\autoref{tab:hyperparameter}, where a range of values indicates, that the
specific hyperparameter will be altered in one of the following experiments. The
exact value can be then found in the corresponding section. To this end, the
topological choice of filter sizes is discussed in \autoref{sec:const_vs_grow}.
The influence of the number of filters of the discriminator and the number of
training samples is shown in \autoref{sec:filter_d}, whereas the effect of the
number of filters of the generator is outlined in \autoref{sec:filter_g}. The
prediction behavior of the generator with respect to varying random input
vectors is analyzed in \autoref{sec:exp1_input}.

\textit{Remark:} The microstructure considered in this section can easily be
generated without the usage of neural networks. Because of its simplicity, it is
perfectly well suited for the benchmark tests carried out. Furthermore, the well
defined shape of the inclusions simplifies qualitative judgement of the
generated microstructures. Interestingly, this seemingly trivial benchmark poses
a significant challenge to GANs, because it has to generate sharp phase
contrasts and perfectly spherical geometries from random noise. This was also
reported in \cite{mosser2017reconstruction}.

\begin{table}[htb] 
    \centering 
    \begin{tabular}{@{}llll@{}} 
        \toprule 
        Hyperparameter & Occurrence & Generator & Discriminator \\
        \midrule 
        Microstructure edge length $\xi_i$ & \autoref{eq:microstructure} & 32-64
        & 32-64 \\
        Batch size $n_B$ & \autoref{sec:example1} & 8 & 8 \\
        Learning rate $\alpha$ & \autoref{sec:example1} & $5 \times 10^{-5}$ & $1
        \times 10^{-4}$ \\
        Nadam moment $\beta_1$ & \autoref{sec:example1} & $0.9$ & $0.9$ \\
        Nadam moment $\beta_2$ & \autoref{sec:example1} & $0.999$ & $0.999$ \\
        LeakyReLU $\gamma$ & \autoref{eq:leakyrelu} & 0.2 & 0.2 \\
        Gradient penalty factor $\lambda$ & \autoref{eq:loss_d} & -- & 10 \\
        No. layers $n_L$ & \autoref{sec:mygan} & 5-6 & 5-6 \\
        Kernel size $n_k$ & \autoref{sec:mygan} & 3 & 3 \\
        Initialization of $\bm{\theta}$ & \autoref{eq:parameters} & orthogonal
        & orthogonal \\
        Input random vector $\bm{z}$ & \autoref{eq:random_input} &
        $\bm{\mathcal{N}}(\bm{0}, \bm{1})$ & -- \\
        Dim. random vector $d_Z$ & \autoref{eq:generator} & 128 & -- \\
        Dim. mapping network  $d_w$ & \autoref{eq:generator} & 128 & -- \\
        No. layers mapping network $n_{Lw}$ & \autoref{eq:ann_map} & 8 & -- \\
        No. filters $n_f$ & \autoref{sec:ann} & 8-64 & 8-64 \\
        \bottomrule 
    \end{tabular} 
    \caption{\textbf{Hyperparameter:} Hyperparameter used in all experiments, if
        not indicated otherwise, for the generator $\mathcal{G}$ from
        \autoref{eq:generator} and the discriminator $\mathcal{D}$ from
    \autoref{eq:discriminator}.}
    \label{tab:hyperparameter} 
\end{table}

\subsection{Filter topology for the discriminator: constant versus growing} 
\label{sec:const_vs_grow} \noindent
First, two topology variants for the discriminator $\mathcal{D}$ from
\autoref{eq:discriminator} are investigated with respect to their performance in
the evaluation metrics from \autoref{sec:metrics} as well as the number of
parameters $\bm{\theta}$ from \autoref{eq:parameters}. The generator topology
described in \autoref{sec:topology_generator} is fixed for both discriminator
topologies and works with a decreasing number of filters to map the low
dimensional random input vector $\bm{z}$ to the image dimension, as shown in
\autoref{fig:topology}. Here, the
number of filters of the last transpose convolutional block is kept at
$n_f^{\mathcal{G}} = 32$. 

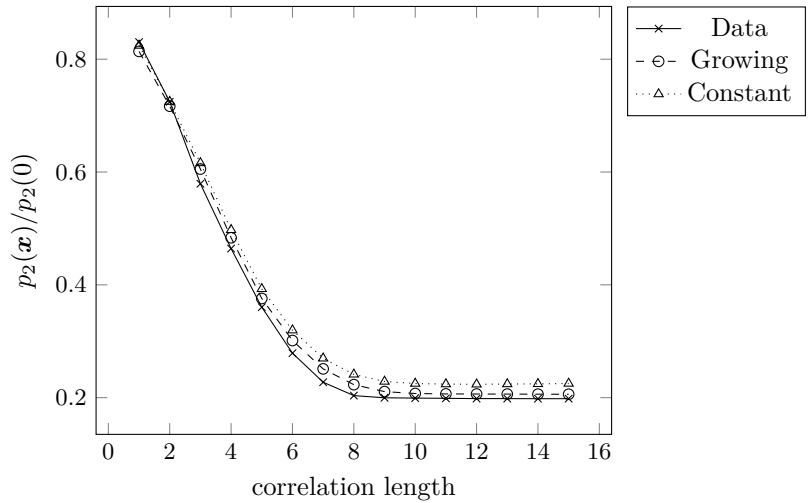
\begin{figure}[htb]
    \centering
    \begin{tikzpicture}
        \begin{axis}[ 
                xlabel={correlation length},
                ylabel={${{p}_2} (\bm{x}) / p_2 (0)$},
                legend pos=outer north east,
            ] 
            \addplot[color=black, mark=x, mark options=solid] 
            coordinates {
                (1, 0.83032143) 
                (2, 0.72456545) 
                (3, 0.57917017) 
                (4, 0.46433184) 
                (5, 0.3604266 ) 
                (6, 0.27888402)
                (7, 0.22740662) 
                (8, 0.20367263) 
                (9, 0.19981   ) 
                (10, 0.19926229) 
                (11, 0.19882788) 
                (12, 0.19836581)
                (13, 0.19833472) 
                (14, 0.19816366) 
                (15, 0.1982123 )
            };
            \addlegendentry{Data}

            \addplot[color=black, dashed, mark=o, mark options=solid] 
            coordinates {
                (1, 0.81380564) 
                (2, 0.71670413) 
                (3, 0.6052457 ) 
                (4, 0.48347718) 
                (5, 0.3757864 ) 
                (6, 0.30130422)
                (7, 0.25103572) 
                (8, 0.22308269) 
                (9, 0.21074048) 
                (10, 0.2073735 ) 
                (11, 0.20665504) 
                (12, 0.20647587)
                (13, 0.20633988) (14, 0.20618024) 
                (15, 0.20607029)
            };
            \addlegendentry{Growing}

            \addplot[color=black, dotted, mark=triangle, mark options=solid] 
            coordinates {
                (1, 0.824239  ) 
                (2, 0.72467893) 
                (3, 0.61599517) 
                (4, 0.4974681 ) 
                (5, 0.3928506 ) 
                (6, 0.31959885)
                (7, 0.26974002) 
                (8, 0.24130592) 
                (9, 0.22867161) 
                (10, 0.22504915) 
                (11, 0.22405624) 
                (12, 0.22369488)
                (13, 0.22418293) 
                (14, 0.22444895) 
                (15, 0.22511852)
            };
            \addlegendentry{Constant}

        \end{axis}
    \end{tikzpicture}
    \caption{\textbf{Example 1: Growing versus constant:} Two-point correlation
    functions for the dataset and the discriminators with growing and constant
filter size from \autoref{sec:const_vs_grow}. Here, a topology with growing
filter size is compared to a topology using constant filter size, where the
total number of parameters $\bm{\theta}$ for both the generator and the
discriminator is kept equal.}
    \label{fig:growing_vs_const} 
\end{figure}

Concerning the discriminator, the number of filters
can be distributed in the same manner as for the generator, i.e. growing or
evenly distributed. While both variants were presented in the literature, e.g.,
\cite{li20193d} for growing and
\cite{mosser2017reconstruction} for constant filters, it is not clear which
variant leads to the best results. Therefore, in this experiment, the total
number of parameters was fixed to lay within in the order of
$\bm{\theta}_{\mathcal{D}} = 3.5 \times 10^6$ while distributing them either in
a growing way or evenly.  Using $n_L = 5$ residual blocks as defined in
\autoref{sec:topology_discriminator} for both topologies, this results in
$n_f^{\mathcal{D}} = 16$ filters for the first layer of the growing
topology and $n_f^{\mathcal{D}} = 120$ filters for the constant topology. Both
the generator and
discriminator are trained for $n_{\text{iter}} = 1 \times 10^5$ iterations. The
number of training
samples was chosen as $1 \times 10^4$. The resulting metrics for $n_s = 1 \times
10^3$ different samples can be seen in \autoref{tab:filters}.  As reported in
\cite{sudret2017surrogate}, this number of samples is sufficient for the mean of
the quantities of interest. The relative error, as defined in
\autoref{eq:error},
of the 1-point probability function or inclusion volume fraction defined in
\autoref{eq:1point} is $\mathcal{E}_{{p}_1} = 4.660 \times 10^{-2}$ for the
growing discriminator topology and $\mathcal{E}_{{p}_1} = 1.378 \times
10^{-1}$ for the constant discriminator topology. The relative error of the
2-point probability function defined in \autoref{eq:2point} is
$\mathcal{E}_{{p}_2} = 2.786 \times 10^{-2}$ for the growing discriminator
topology and $\mathcal{E}_{{p}_2} = 6.892 \times 10^{-2}$ for the constant
discriminator topology. The 2-point probability distributions of both
topologies are illustrated in \autoref{fig:growing_vs_const}. The relative error
of the homogenized $\bar{\mathbb{C}}_{11}$ component as defined in
\autoref{sec:homogenization} is $\mathcal{E}_{\bar{\mathbb{C}}_{11}} = 7.266
\times 10^{-3}$ for the growing discriminator topology and
$\mathcal{E}_{\bar{\mathbb{C}}_{11}} = 6.399 \times 10^{-2}$ for the constant
discriminator topology.  

Besides leading to worse results, the constant filter topology takes with 3h
significantly longer time to train, compared to the growing topology with 1h.
The reason for this is the convolution operation with respect to the larger
number of filters in the first convolutional block for the constant layout
$n_f^{\mathcal{D}} = 120$ versus the growing layout $n_f^{\mathcal{D}} = 16$. In
the first block, the spatial dimensions of the are high compared to the later
blocks. Here, the convolution operation is significantly more expansive. For the
rest of this paper, the growing topology is used for the discriminator and a
decreasing topology for the generator, where for the latter the same reasoning
applies. 

\begin{table}[htb] 
    \centering 
    \begin{tabular}{@{}llll@{}} 
        \toprule 
        Topology & 
        $\mathcal{E}_{{p}_1}$ & 
        $\mathcal{E}_{{p}_2}$ & 
        $\mathcal{E}_{\bar{\mathbb{C}}_{11}}$ \\
        \midrule 
        Growing 
        & $4.660 \times 10^{-2}$ 
        & $2.786 \times 10^{-2}$ 
        & $7.266 \times 10^{-3}$ 
        \\
        Constant 
        & $1.378 \times 10^{-1}$
        & $6.892 \times 10^{-2}$
        & $6.399 \times 10^{-2}$ 
        \\
        \bottomrule 
    \end{tabular} 
    \caption{\textbf{Example 1:} Deviation from a reference solution using
        different discriminator topologies as described in
        \autoref{sec:const_vs_grow}. Here, a topology with growing filter size
        is compared to a topology using constant filter size, where the total
    number of parameters $\bm{\theta}$ for both the generator and the
discriminator is kept equal.}
    \label{tab:filters} 
\end{table}

\subsection{Influence of filter number for the discriminator}
\label{sec:filter_d} \noindent
This section deals with the influence of the number of filters
$n_f^{\mathcal{D}}$ of the discriminator $\mathcal{D}$ on the ability to
generate synthetic microstructures, measured with respect to the first component
of the homogenized elasticity tensor, ${\bar{\mathbb{C}}_{11}}$. To
this end, a fixed generator topology with $n_f^{\mathcal{G}} = 32$ filters in
the last transpose residual convolution block $\mathcal{N}_{\text{ResConvT}}$
from \autoref{eq:block_g} is chosen and a total of $n_L^{\mathcal{G}} = 5$
blocks is used. For the discriminator, the number of filters $n_f^{\mathcal{D}}
= 32$ in
the first residual convolutional block $\mathcal{N}_{\text{ResConv}}$ from
\autoref{eq:block_d}  are varied, such that $n_f^{\mathcal{D}} =
[8, 16, 32, 64]$, where the total number of weights $\bm{\theta}_{\mathcal{D}}$
grows exponentially. The number of convolutional blocks is $n_L^{\mathcal{D}} =
5$ for all discriminators. For $n_f^{\mathcal{D}} = [8, 16, 32, 64]$ and the
given number
of blocks, the number of weights for the discriminator are
$\bm{\theta}_{\mathcal{D}} = [0.9, 3.5, 14.3, 57] \times 10^6$, respectively.
To investigate the effect of the number of training samples $n_s$ of the dataset
$\mathbb{D}$ from \autoref{eq:dataset} on the different discriminator
topologies, all networks are trained on three different amounts of
microstructural realizations, namely $n_s = [10^2, 10^3, 10^4]$. For all
experiments, the discriminator and generator are trained for $n_{\text{iter}} =
1 \times 10^6$ steps each. 

\begin{figure}[htb]
    \centering
    \begin{tikzpicture}
        \begin{axis}[ 
                xlabel={No. of filters in first layer of the discriminator
                $n_f^{\mathcal{D}}$},
                ylabel={$\mathcal{E}_{\bar{\mathbb{C}}_{11}}$},
                legend pos=outer north east,
                xticklabels={
                    {$8$}, 
                    {$16$}, 
                    {$32$}, 
                    {$64$}, 
                },
                xtick={1,...,4},
            ] 
            \addplot[color=black, mark=x, mark options=solid] 
            coordinates {
                (1, 0.017425814643502235)
                (2, 0.01699395291507244)
                (3, 0.00864303670823574)
                (4, 0.006220459006726742)
            };
            \addlegendentry{$n_s = 10^4$}

            \addplot[color=black, dashed, mark=o, mark options=solid] 
            coordinates {
                (1, 0.016788721084594727)
                (2, 0.016602154821157455)
                (3, 0.006214876659214497)
                (4, 0.019452491775155067)
            };
            \addlegendentry{$n_s = 10^3$}

            \addplot[color=black, dotted, mark=triangle, mark options=solid] 
            coordinates {
                (1, 0.0156913623213768)
                (2, 0.030564816668629646)
                (3, 0.04359371215105057)
                (4, 0.04184252768754959)
            };
            \addlegendentry{$n_s = 10^2$}

        \end{axis}
    \end{tikzpicture}
    \caption{\textbf{Example 1: Filter discriminator:} Relative error of the
        homogenized elasticity tensor component $\bar{\mathbb{C}}_{11}$ with
        respect to
    the number of starting filters $n_f^{\mathcal{D}}$ of the discriminator and
different number of samples $n_s$ from \autoref{sec:filter_d}. The
hyperparameters of the generator are fixed.}
    \label{fig:filter_d_c11} 
\end{figure}
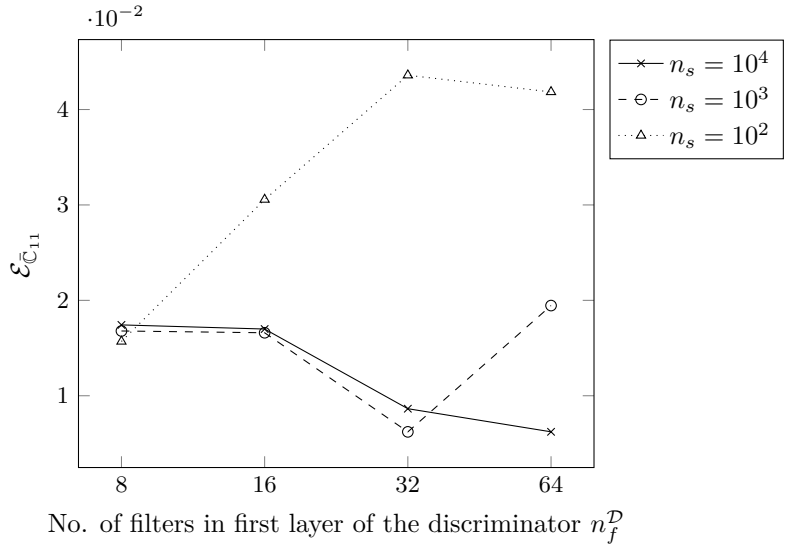

The resulting error of the first component of the elasticity tensor
$\mathcal{E}_{\bar{\mathbb{C}}_{11}}$ is shown in \autoref{fig:filter_d_c11}. A
clear effect of both the number of filters as well as the number of training
samples can be observed. For the smallest amount of training samples, $n_s = 1
\times 10^2$, the error increases for higher filter numbers until
$n_f^{\mathcal{D}} = 32$, where a plateau is reached. Using $n_s = 1 \times
10^3$ samples, a different
behavior is observed. Here the error goes down for increasing filter numbers
until $n_f^{\mathcal{D}} = 32$, from where on the error rises. For the largest
amount of training samples, $n_s = 1 \times 10^4$, the error goes down steadily,
despite of the number of filters. This effect can be explained by an overfitting
behavior of the discriminator, when the capacity, by means of
the total number of weights $\bm{\theta}_{\mathcal{D}}$, is large with respect
to the number of samples $n_s$. Then, the discriminator will always be able to
distinguish
between generated images and training samples and assign large Wasserstein
distances to them, despite the quality of the synthetic images. This gives the
generator poor gradient updates. The significance of this effect towards
practical application is high.  One has to take care of choosing the appropriate
discriminator for the
specific number of training samples at hand to avoid overfitting effects.

\subsection{Influence of filter number for the generator}
\label{sec:filter_g} \noindent
This section deals with the influence of the number of filters
$n_f^{\mathcal{G}}$ of the generator $\mathcal{G}$ on the ability to
generate synthetic microstructures, measured with respect to the first component
of the homogenized elasticity tensor, ${\bar{\mathbb{C}}_{11}}$. To
this end, a fixed discriminator topology with $n_f^{\mathcal{D}} = 64$ filters
in the first residual convolution block $\mathcal{N}_{\text{ResConv}}$ from
\autoref{eq:block_d} is chosen and a total of $n_L^{\mathcal{D}} = 5$ blocks is
used. This is
the best performing topology from \autoref{sec:filter_d}. For the
generator the number of filters $n_f^{\mathcal{G}} = 32$ in the last
transpose residual convolutional block $\mathcal{N}_{\text{ResConvT}}$ from
\autoref{eq:block_g} are varied, such that $n_f^{\mathcal{G}} = [8, 16, 32,
64]$, where the total number of weights $\bm{\theta}_{\mathcal{G}}$ grows
exponentially. The number of transpose convolutional blocks is
$n_L^{\mathcal{G}} = 5$ for all generators. For $n_f^{\mathcal{G}} = [8, 16, 32,
64]$ and the given number of blocks, the number
of weights for the generator are $\bm{\theta}_{\mathcal{G}} = [0.3, 0.6, 1.6,
5.5] \times 10^6$, respectively. The number of training samples $n_s$ of the
dataset $\mathbb{D}$ from \autoref{eq:dataset} is fixed at $n_s = 10^4$ to
prevent overfitting as observed in \autoref{sec:filter_d}. For all experiments,
the discriminator and generator are trained for $n_{\text{iter}} = 1 \times
10^6$ steps each. 

\begin{figure}[htb]
    \centering
    \begin{tikzpicture}
        \begin{axis}[ 
                xlabel={No. of filters in first layer of the generator
                $n_f^{\mathcal{G}}$},
                ylabel={$\mathcal{E}_{\bar{\mathbb{C}}_{11}}$},
                legend pos=outer north east,
                xticklabels={
                    {$8$}, 
                    {$16$}, 
                    {$32$}, 
                    {$64$}, 
                },
                xtick={1,...,4},
            ] 
            \addplot[color=black, mark=x, mark options=solid] 
            coordinates {
                (1, 0.012128935195505619) 
                (2, 0.0178562980145216)
                (3, 0.006220459006726742) 
                (4, 0.006639049854129553) 
            };
            \addlegendentry{$n_{\text{iter}} = 1 \times 10^6$}

            \addplot[color=black, dashed, mark=o, mark options=solid] 
            coordinates {
                (3, 0.006220459006726742) 
                (4, 0.003907473757863045) 
            };
            \addlegendentry{$n_{\text{iter}} = 2 \times 10^6$}

        \end{axis}
    \end{tikzpicture}
    \caption{\textbf{Example 1: Filter generator:} Relative error of the
        homogenized elasticity tensor component $\bar{\mathbb{C}}_{11}$ with
        respect to the number of starting filters $n_f^{\mathcal{G}}$ of the
        generator for
        $n_{\text{iter}} = 10^6$ iterations from \autoref{sec:filter_g}. The
        hyperparameters of the discriminator are fixed. The generator with
    $n_f=64$ filters did not converge within the maximum number of iterations.
Therefore, an additional for $n_{\text{iter}} = 2 \times 10^6$ iterations was
carried out and reported.}
        \label{fig:filter_g} 
\end{figure}
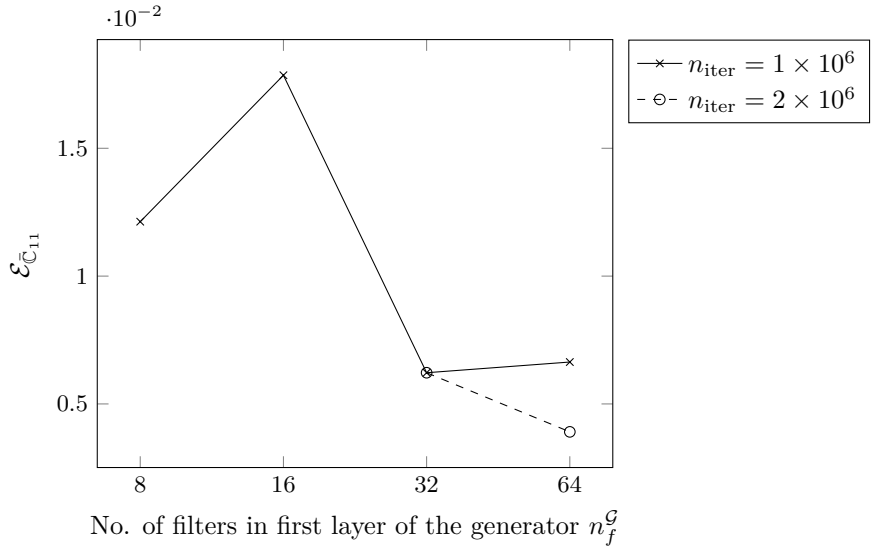

The resulting error of the first component of the elasticity tensor
$\mathcal{E}_{\bar{\mathbb{C}}_{11}}$ is shown in \autoref{fig:filter_g}. As no
convergence for the largest topology, $n_f^{\mathcal{G}} = 64$ could be
observed, for that specific number of filters a re-run of the simulation for
$n_{\text{iter}} = 2 \times 10^6$ steps was carried out. Besides slower
convergence, a trend towards lower errors for larger generators, in the sense of
parameters
$\bm{\theta}_{\mathcal{G}}$ and filter size $n_f^{\mathcal{G}}$, can be
observed. This indicates the usage of the largest possible generator topology
available for the given computational power to achieve the lowest error.

\subsection{Influence of random input vector}
\label{sec:exp1_input} 
\begin{figure}[htb]
    \centering
    \begin{subfigure}[b]{0.3\textwidth}
        \centering
        \includegraphics[width=1.0\textwidth]{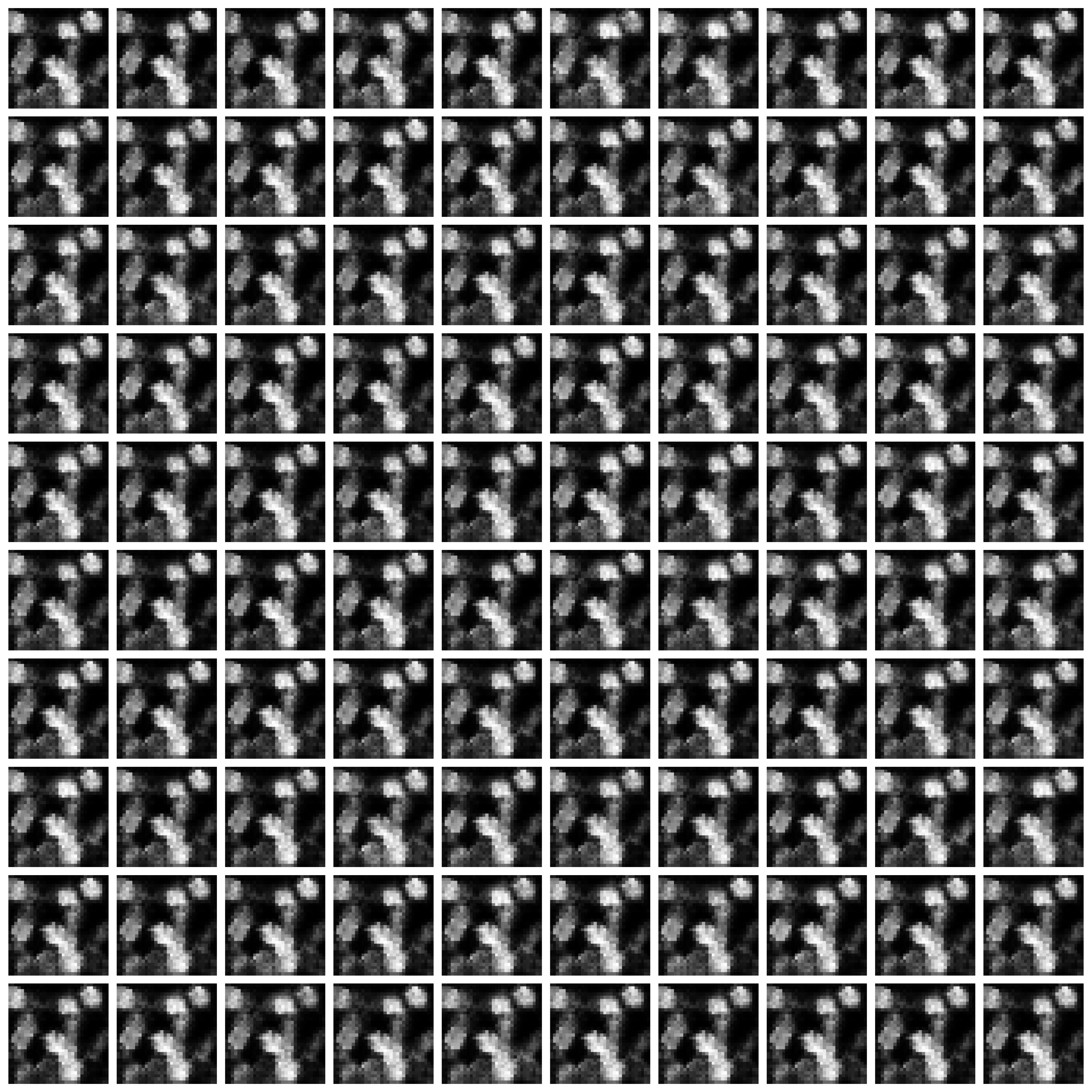}
        \caption{$n_f^{\mathcal{D}} = 64, \sigma_{\text{pred}} = 0.1$}
        \label{fig:64_01}
    \end{subfigure}
    \hfill
    \begin{subfigure}[b]{0.3\textwidth}
        \centering
        \includegraphics[width=1.0\textwidth]{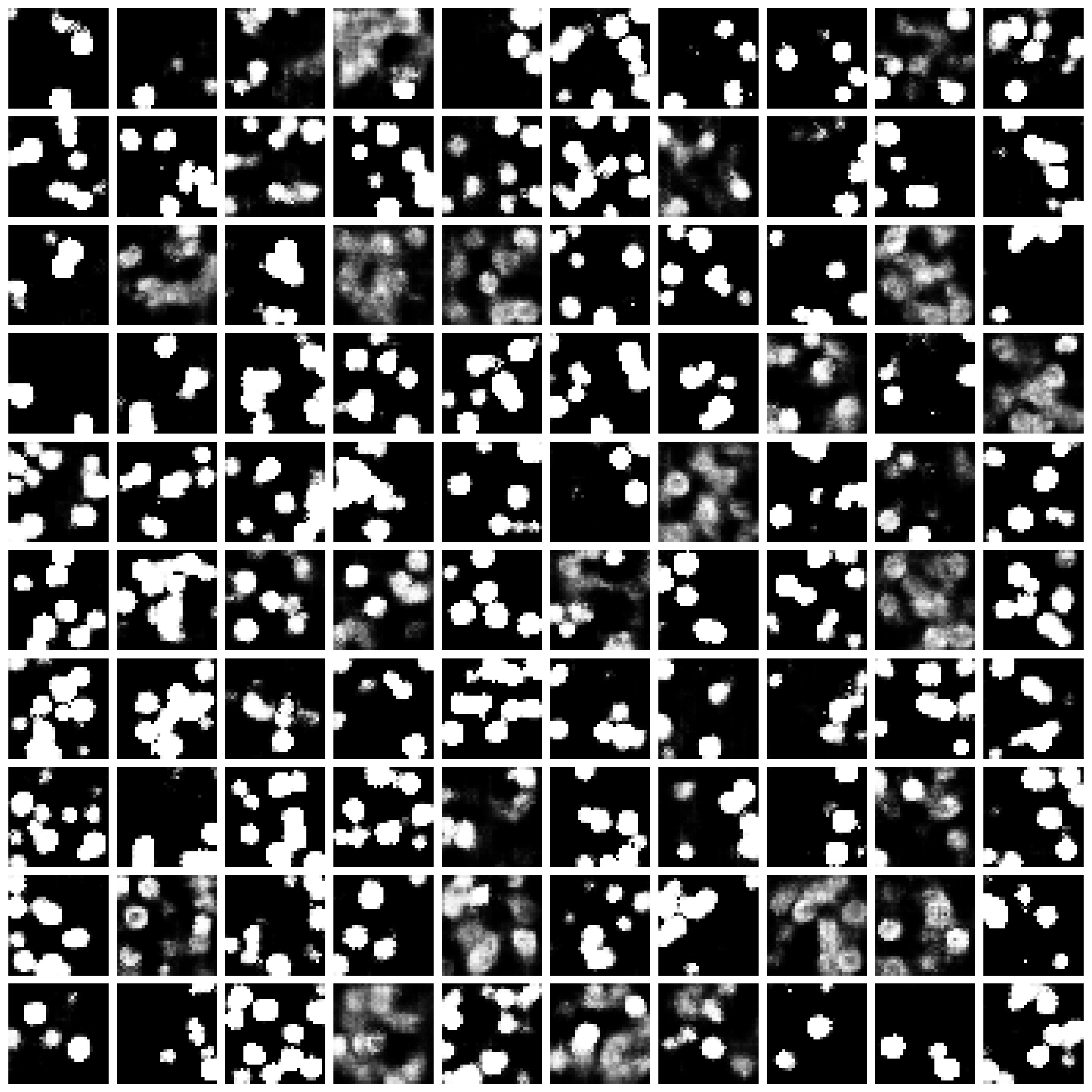}
        \caption{$n_f^{\mathcal{D}} = 64, \sigma_{\text{pred}} = 1.0$}
        \label{fig:64_1}
    \end{subfigure}
    \hfill
    \begin{subfigure}[b]{0.3\textwidth}
        \centering
        \includegraphics[width=1.0\textwidth]{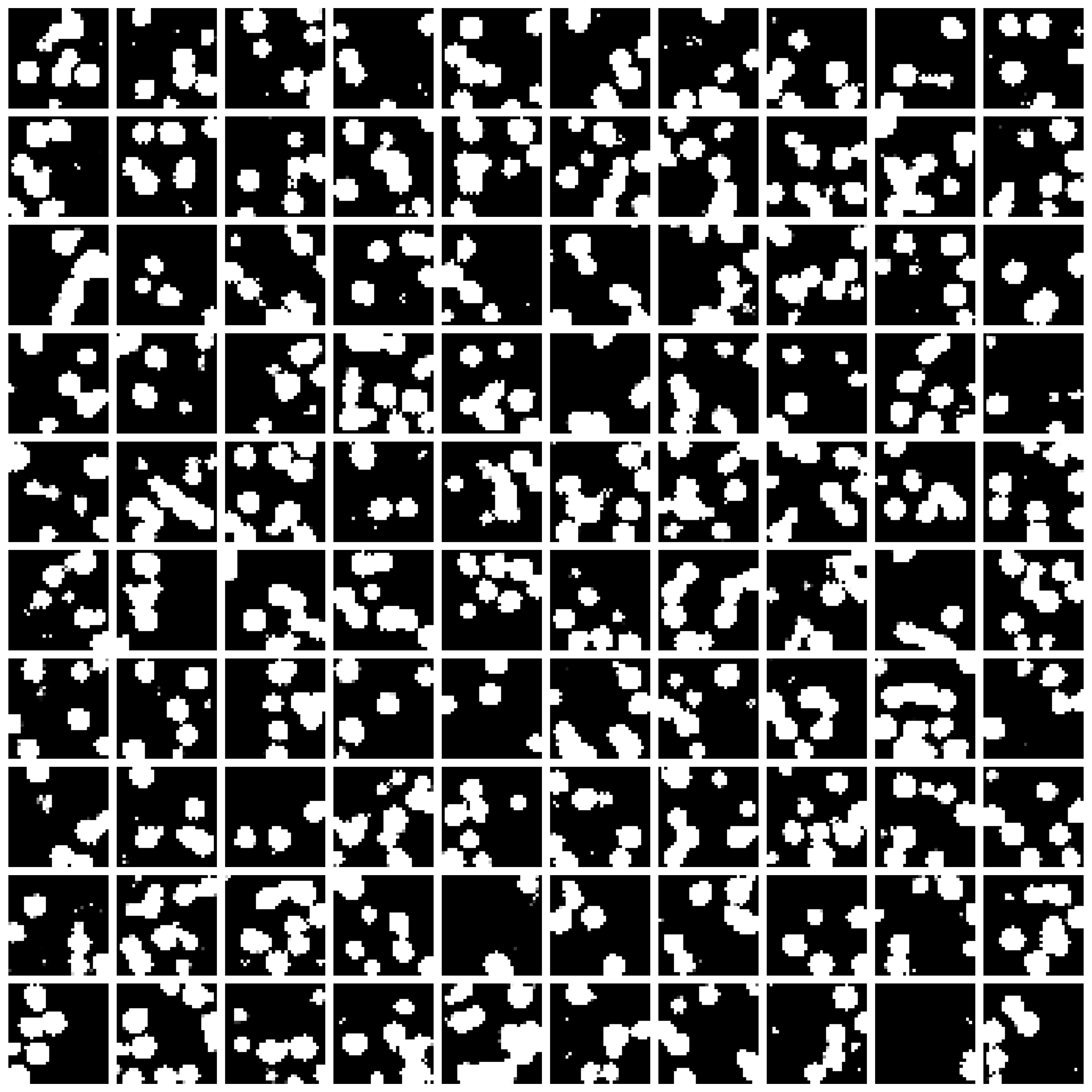}
        \caption{$n_f^{\mathcal{D}} = 64, \sigma_{\text{pred}} = 10.0$}
        \label{fig:64_10}
    \end{subfigure}

    \begin{subfigure}[b]{0.3\textwidth}
        \centering
        \includegraphics[width=1.0\textwidth]{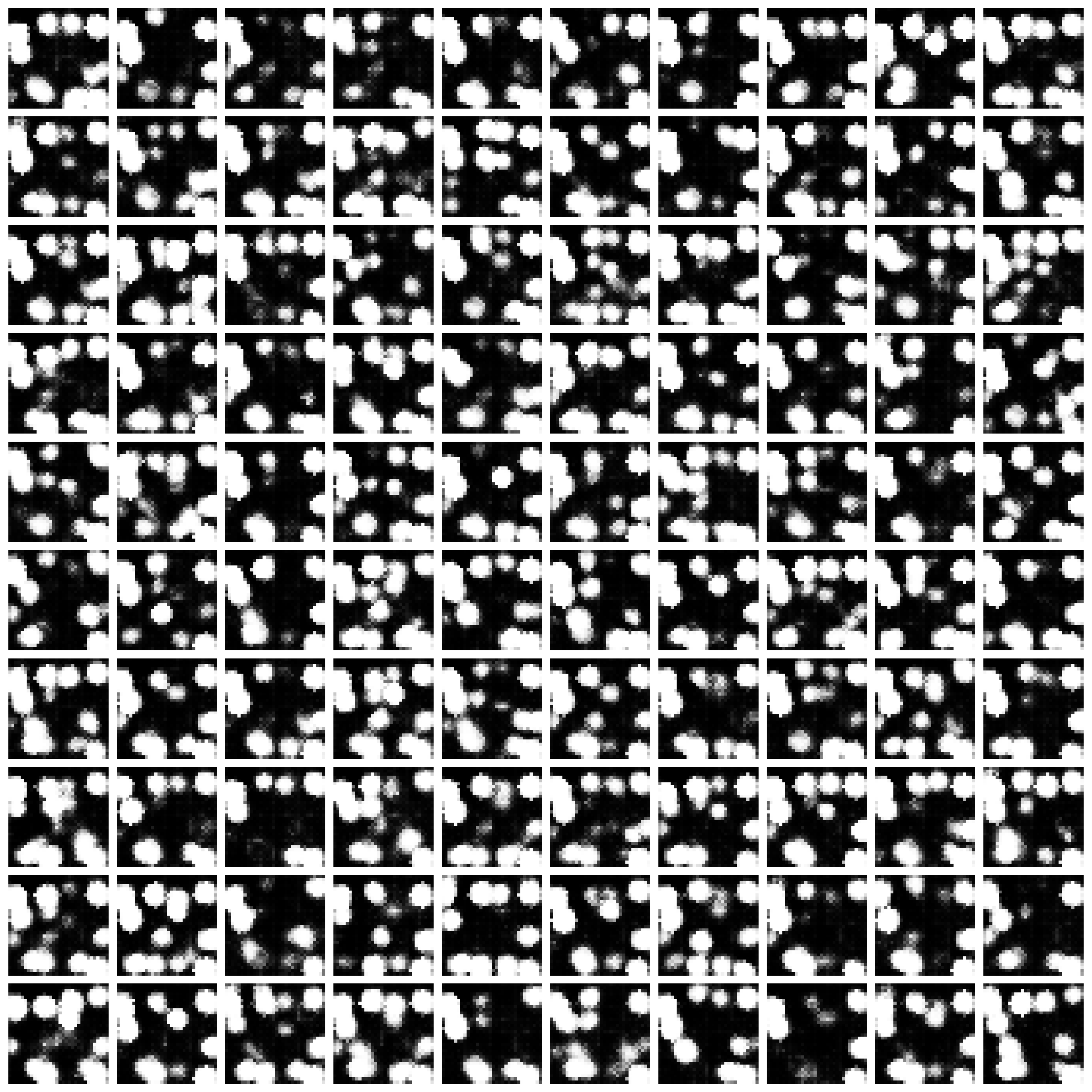}
        \caption{$n_f^{\mathcal{D}} = 32, \sigma_{\text{pred}} = 0.1$}
        \label{fig:32_01}
    \end{subfigure}
    \hfill
    \begin{subfigure}[b]{0.3\textwidth}
        \centering
        \includegraphics[width=1.0\textwidth]{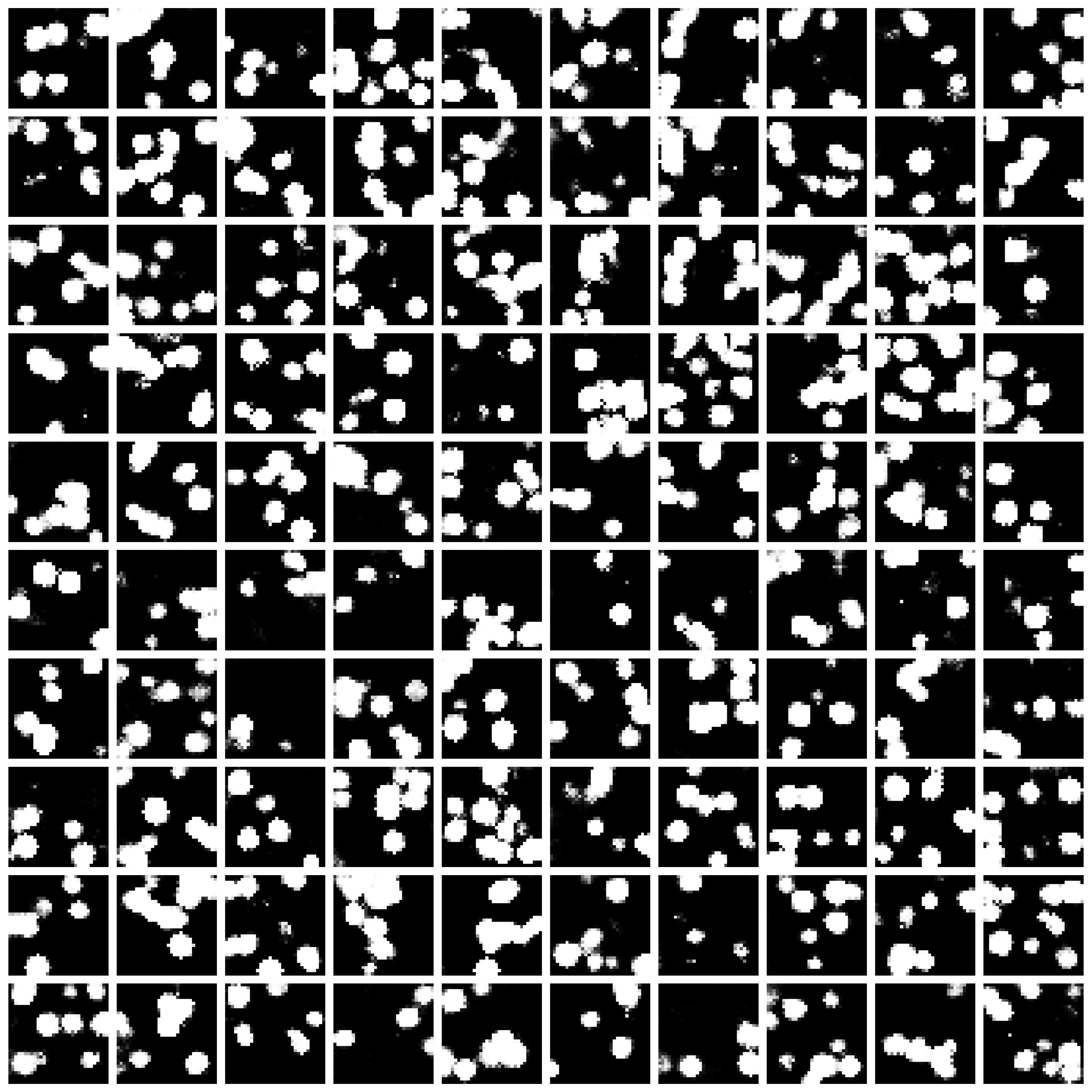}
        \caption{$n_f^{\mathcal{D}} = 32, \sigma_{\text{pred}} = 1.0$}
        \label{fig:32_1}
    \end{subfigure}
    \hfill
    \begin{subfigure}[b]{0.3\textwidth}
        \centering
        \includegraphics[width=1.0\textwidth]{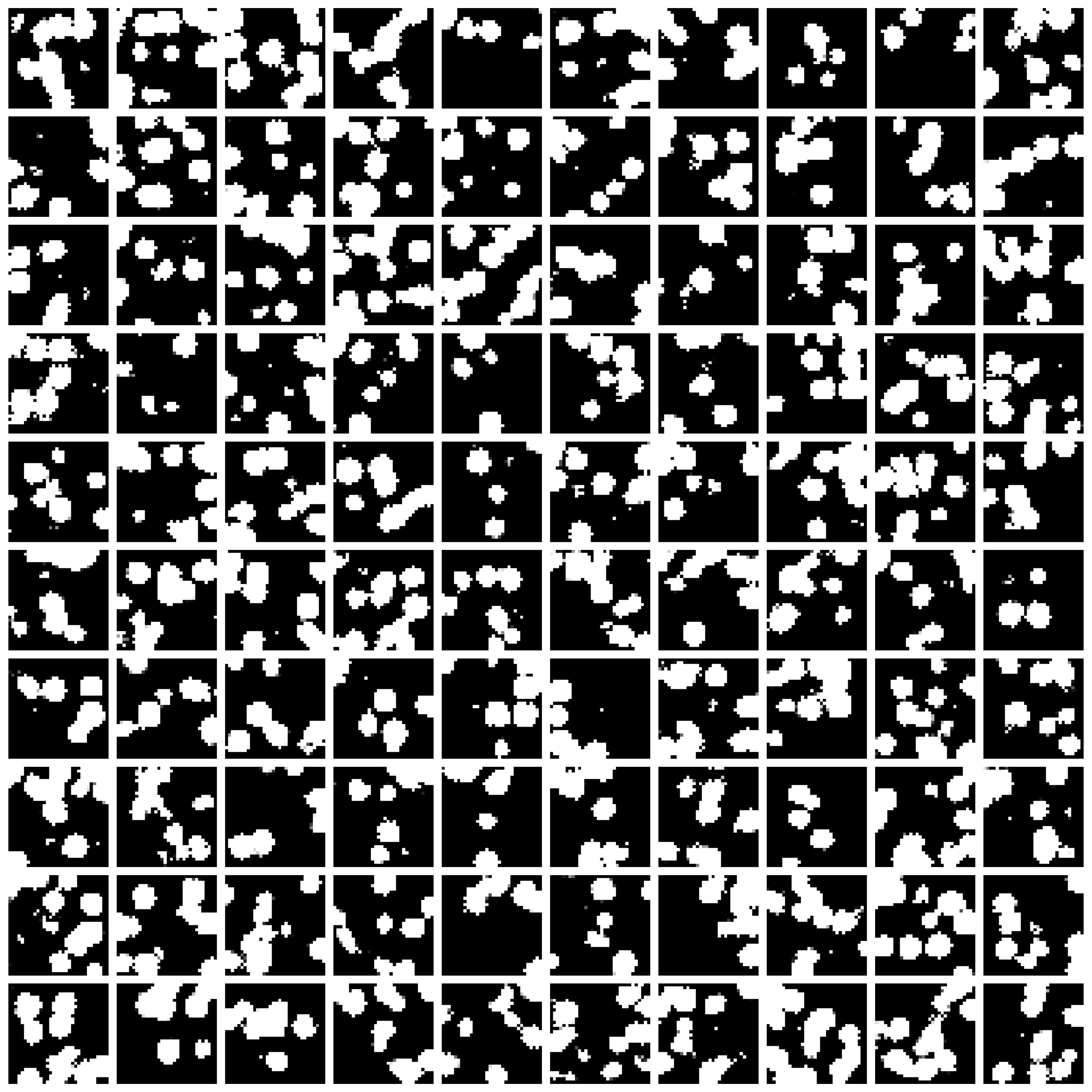}
        \caption{$n_f^{\mathcal{D}} = 32, \sigma_{\text{pred}} = 10.0$}
        \label{fig:32_10}
    \end{subfigure}

    \begin{subfigure}[b]{0.3\textwidth}
        \centering
        \includegraphics[width=1.0\textwidth]{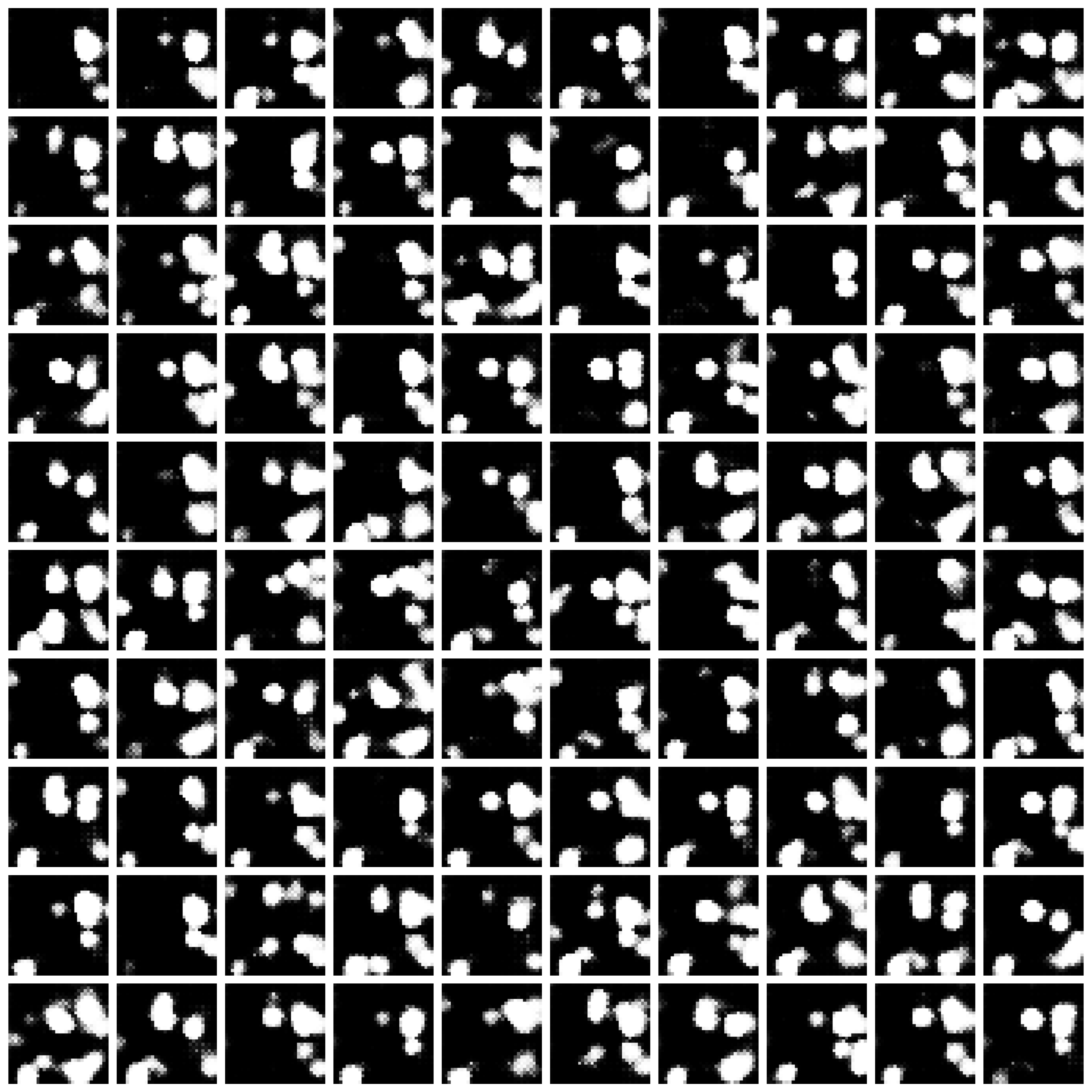}
        \caption{$n_f^{\mathcal{D}} = 16, \sigma_{\text{pred}} = 0.1$}
        \label{fig:16_01}
    \end{subfigure}
    \hfill
    \begin{subfigure}[b]{0.3\textwidth}
        \centering
        \includegraphics[width=1.0\textwidth]{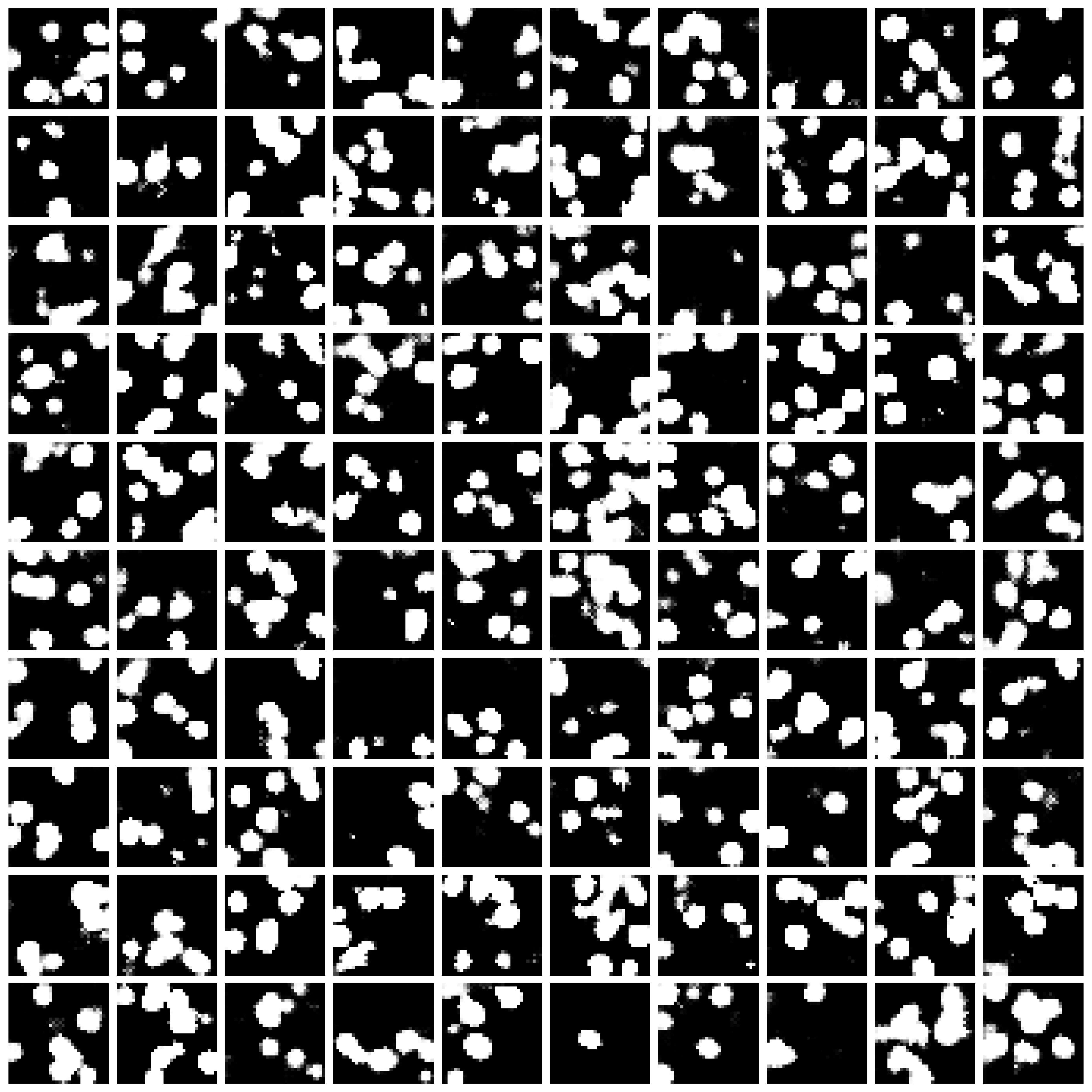}
        \caption{$n_f^{\mathcal{D}} = 16, \sigma_{\text{pred}} = 1.0$}
        \label{fig:16_1}
    \end{subfigure}
    \hfill
    \begin{subfigure}[b]{0.3\textwidth}
        \centering
        \includegraphics[width=1.0\textwidth]{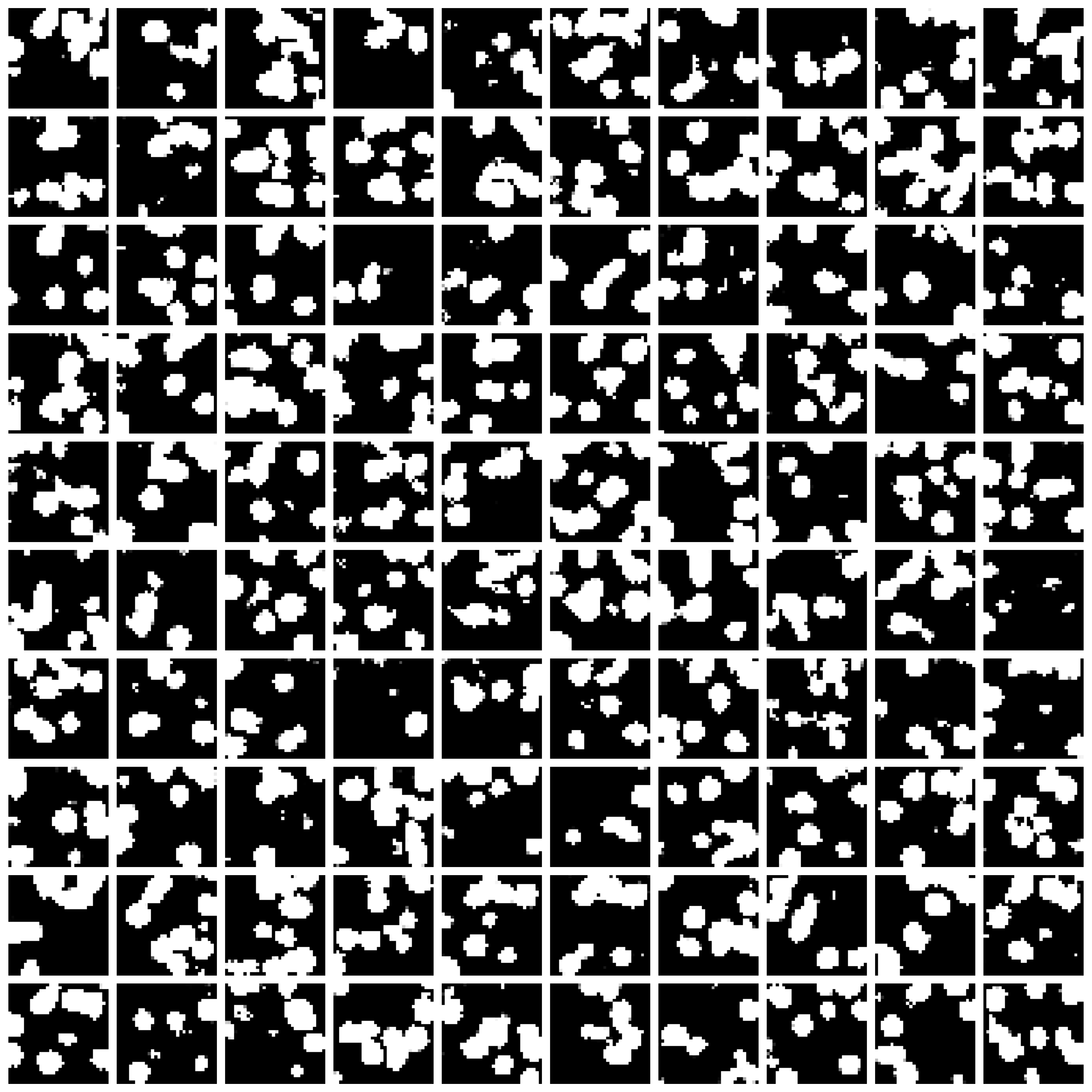}
        \caption{$n_f^{\mathcal{D}} = 16, \sigma_{\text{pred}} = 10.0$}
        \label{fig:16_10}
    \end{subfigure}

    \begin{subfigure}[b]{0.3\textwidth}
        \centering
        \includegraphics[width=1.0\textwidth]{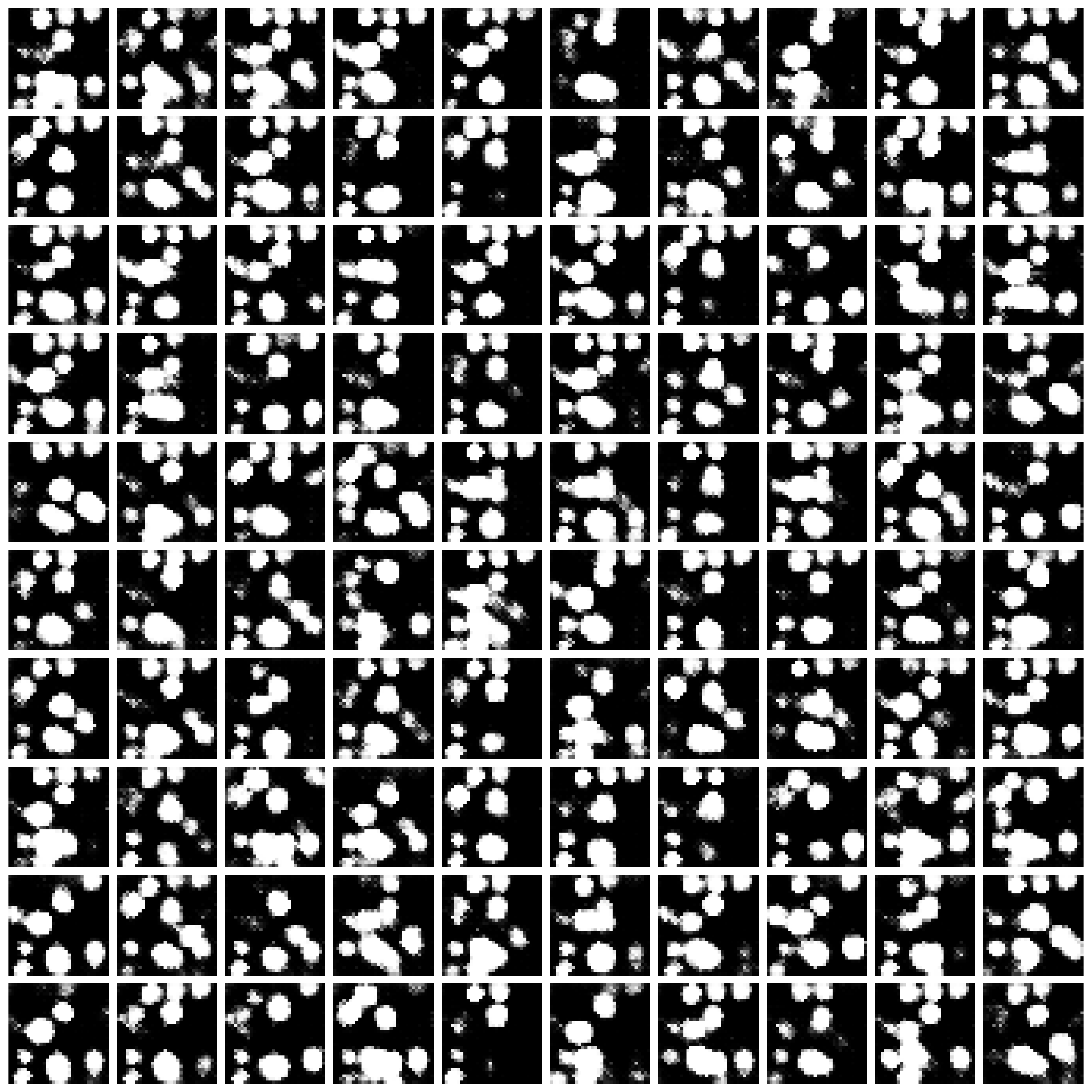}
        \caption{$n_f^{\mathcal{D}} = 8, \sigma_{\text{pred}} = 0.1$}
        \label{fig:8_01}
    \end{subfigure}
    \hfill
    \begin{subfigure}[b]{0.3\textwidth}
        \centering
        \includegraphics[width=1.0\textwidth]{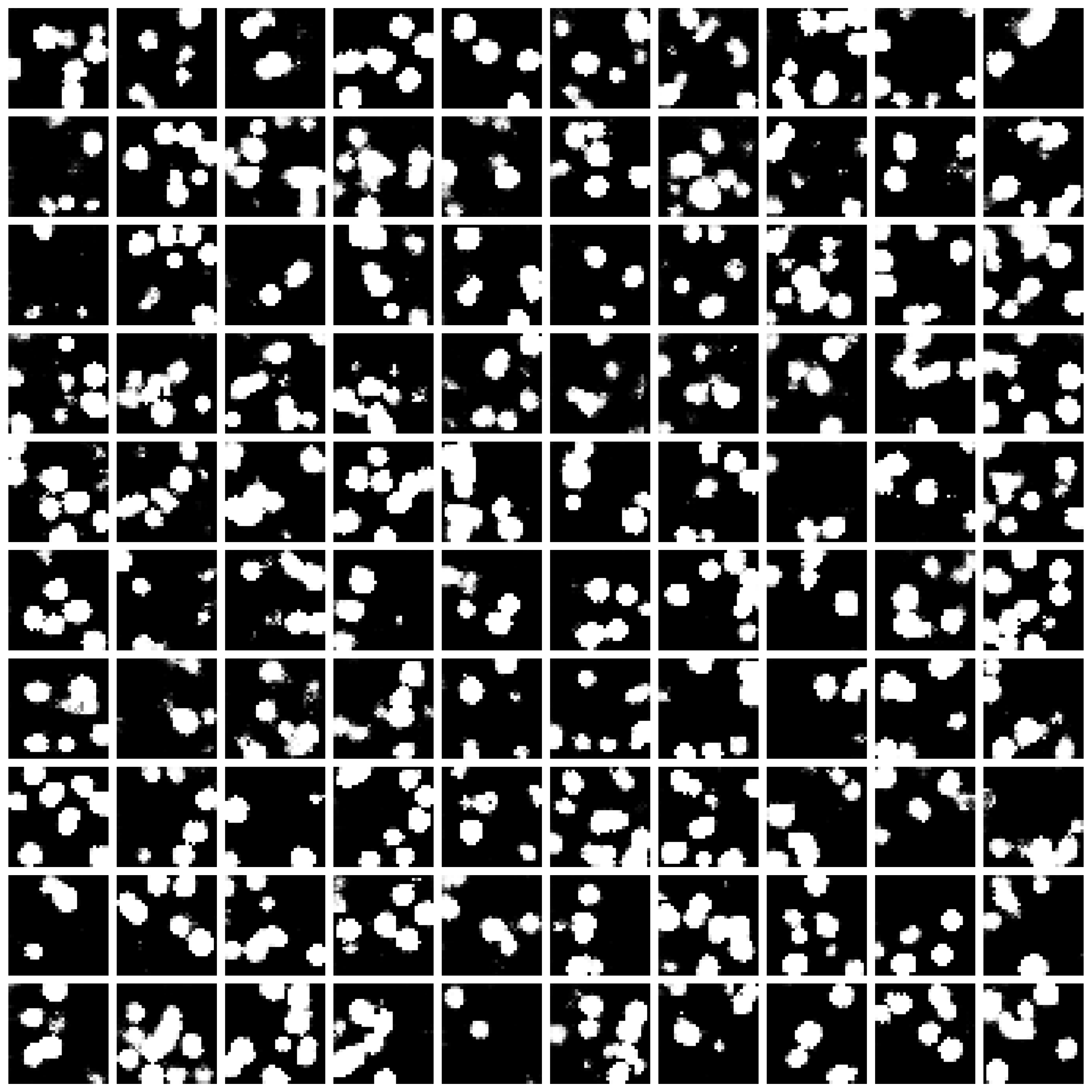}
        \caption{$n_f^{\mathcal{D}} = 8, \sigma_{\text{pred}} = 1.0$}
        \label{fig:8_1}
    \end{subfigure}
    \hfill
    \begin{subfigure}[b]{0.3\textwidth}
        \centering
        \includegraphics[width=1.0\textwidth]{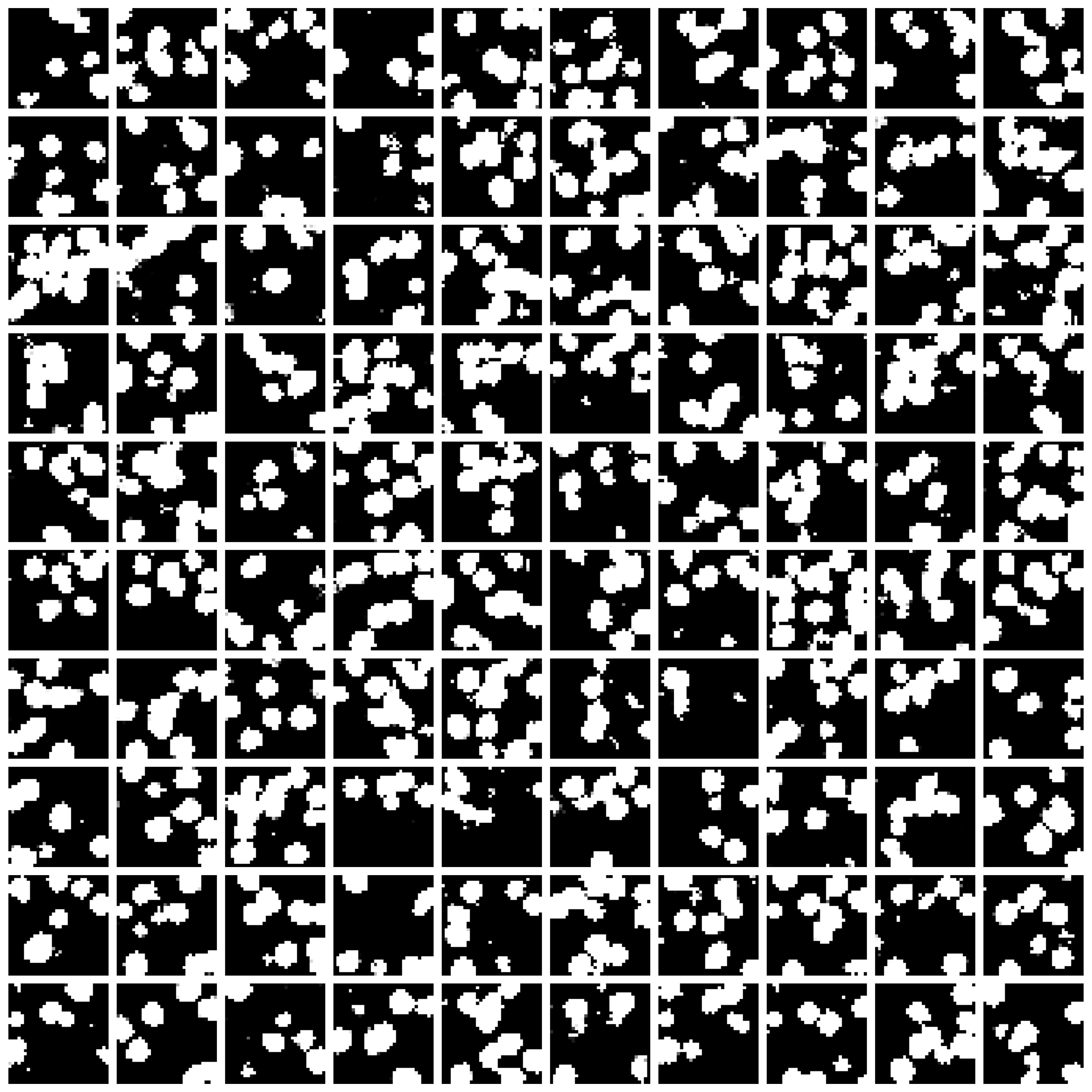}
        \caption{$n_f^{\mathcal{D}} = 8, \sigma_{\text{pred}} = 10.0$}
        \label{fig:8_10}
    \end{subfigure}

    \caption{\textbf{Example 1: Random input vector:} Generated images of
        spherical inclusions from \autoref{sec:example1} for different
        discriminator filter sizes $n_f^{\mathcal{D}}$ used during training from
        \autoref{eq:discriminator}. The generator is kept fixed with
        $n_f^{\mathcal{G}} = 32$. Varying standard deviations of the random
        input vector $\bm{z}_{\text{pred}}$ from \autoref{eq:random_input} are
        chosen. (a) $\sigma_{\text{pred}} = 0.1$, (b) $\sigma_{\text{pred}} =
    1.0$, (c) $\sigma_{\text{pred}} = 10.0$.  Details can be found in
\autoref{sec:exp1_input}}
    \label{fig:stdev}
\end{figure}

\noindent
Another interesting effect is observed at prediction time. To this end, in this
section two random input vectors
are distinguished explicitly, namely the \textit{training random input vector}
$\bm{z}_{\text{train}}$ and the \textit{prediction random input vector}
$\bm{z}_{\text{pred}}$. During training, the generator from
\autoref{eq:generator} takes a normally distributed random vector
$\bm{z}_{\text{train}}$ as input. Usually, the mean $\mu$ and standard deviation
$\sigma$ from \autoref{eq:random_input} are chosen as $\mu_{\text{train}} = 0$
and $\sigma_{\text{train}} = 1$, which results in the best training dynamics.
Details can be found in standard texts such as \cite{aggarwal2018neural} or
\cite{goodfellow2016deep}. In the experiments carried out in this section it was
found that during prediction, the quality and variety of the generated
microstructures strongly depends on the standard deviation
$\sigma_{\text{pred}}$ of the random input $\bm{z}_{\text{pred}}$ and the
discriminator capacity, in terms of parameters $\bm{\theta}_{\mathcal{D}}$, used
during training. To illustrate this effect, we plot predictions of the same
generator $\mathcal{G}$ from \autoref{eq:generator} trained with different
discriminators $\mathcal{D}$ from \autoref{eq:discriminator} with varying
standard deviations $\sigma_{\text{pred}} = [0.1, 1.0, 10.0]$ in
$\bm{z}_{\text{pred}}$. Here, a generator
with $n_f^{\mathcal{G}} = 32$ filters was trained with discriminators using a
filter size of $n_f^{\mathcal{D}} = [8, 16, 32, 64]$. The remaining
hyperparameters are chosen equally to that reported in \autoref{sec:filter_d}.
The resulting microstructures are shown in \autoref{fig:stdev}. 

It can be observed that the observed variety and quality of the generated images
is lower for larger discriminators with a large number of filters
$n_f^{\mathcal{D}}$
using a small standard deviation $\sigma_{\text{pred}}$. The larger the standard
deviation is chosen, the higher the variety of the generated images.
Furthermore, the largest discriminator with $n_f^{\mathcal{D}} = 64$ shows
significant disturbance in the images for small and medium standard deviations
of $\sigma_{\text{pred}} = [0.1, 1.0]$. For smaller discriminators, this effect
weakens the smaller the number of filters gets. For the smallest discriminator
with $n_f^{\mathcal{D}} = 8$, for all investigated standard deviations, varying
microstructures without defects could be generated. For the largest standard
deviation, all networks produced high quality and diverse microstructures.

This effect is different from the well known \textit{truncation trick} 
reported in \cite{marchesi2017megapixel}, where a normal random input variable
was replaced by a truncated normal random variable to reduce variance. Therein,
the reduced variance led to higher image quality but smaller variation.

Concerning practical implications, care should be taken if large networks are
used and only a small number of training samples is available. Then, a larger
standard deviation than during training should be chosen during prediction time. 

\FloatBarrier
\subsection{Best model}
\label{sec:best} 
\begin{figure}[htb]
    \centering
    \begin{tikzpicture}
        \begin{axis}[ 
                xlabel={correlation length},
                ylabel={$p_2 (\bm{x}) / p_2 (0)$},
                legend pos=outer north east,
            ] 
            \addplot[color=black, mark=x, mark options=solid] 
            coordinates {
                (1, 0.83032143) 
                (2, 0.72456545) 
                (3, 0.57917017) 
                (4, 0.46433184) 
                (5, 0.3604266 ) 
                (6, 0.27888402)
                (7, 0.22740662) 
                (8, 0.20367263) 
                (9, 0.19981   ) 
                (10, 0.19926229) 
                (11, 0.19882788) 
                (12, 0.19836581)
                (13, 0.19833472) 
                (14, 0.19816366) 
                (15, 0.1982123 )
            };
            \addlegendentry{Data}

            \addplot[color=black, dashed, mark=o, mark options=solid] 
            coordinates {
                (1, 0.8248339)  
                (2, 0.7238541)  
                (3, 0.58455396) 
                (4, 0.47420758) 
                (5, 0.3738202)  
                (6, 0.29368645)
                (7, 0.24216455) 
                (8, 0.21635248) 
                (9, 0.2088925)  
                (10, 0.20778644) 
                (11, 0.20697016) 
                (12, 0.20601569)
                (13, 0.20555417) 
                (14, 0.2050343)  
                (15, 0.20482783)
            };
            \addlegendentry{GAN}
        \end{axis}
    \end{tikzpicture}
    \caption{\textbf{Example 1: Best model.} Two-point correlation
    functions for the dataset and a generator with $n_f^{\mathcal{G}} = 64$
        filters trained by a discriminator with $n_f^{\mathcal{D}} = 64$
    filters as described in \autoref{sec:best}.}
    \label{fig:best2p} 
\end{figure}
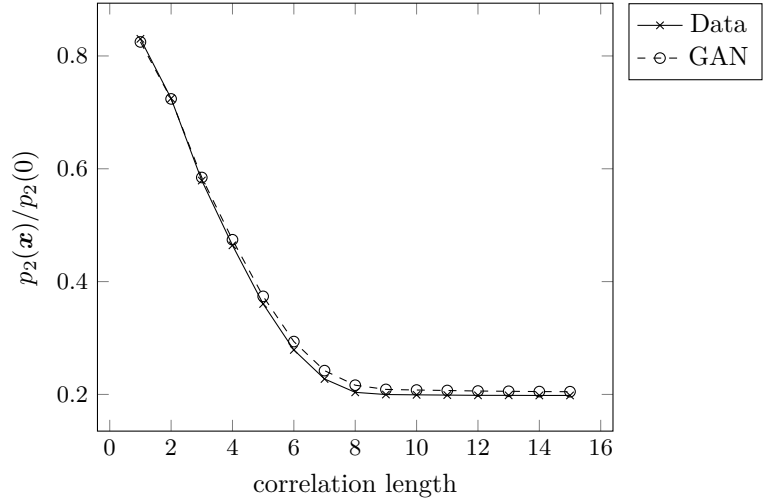

\begin{figure}[htb]
    \centering
    \begin{subfigure}[b]{0.45\textwidth}
        \centering
        \includegraphics[width=1.0\textwidth]{./microstructure_real.pdf}
        \caption{Real}
        \label{fig:data}
    \end{subfigure}
    \hfill
    \begin{subfigure}[b]{0.45\textwidth}
        \centering
        \includegraphics[width=1.0\textwidth]{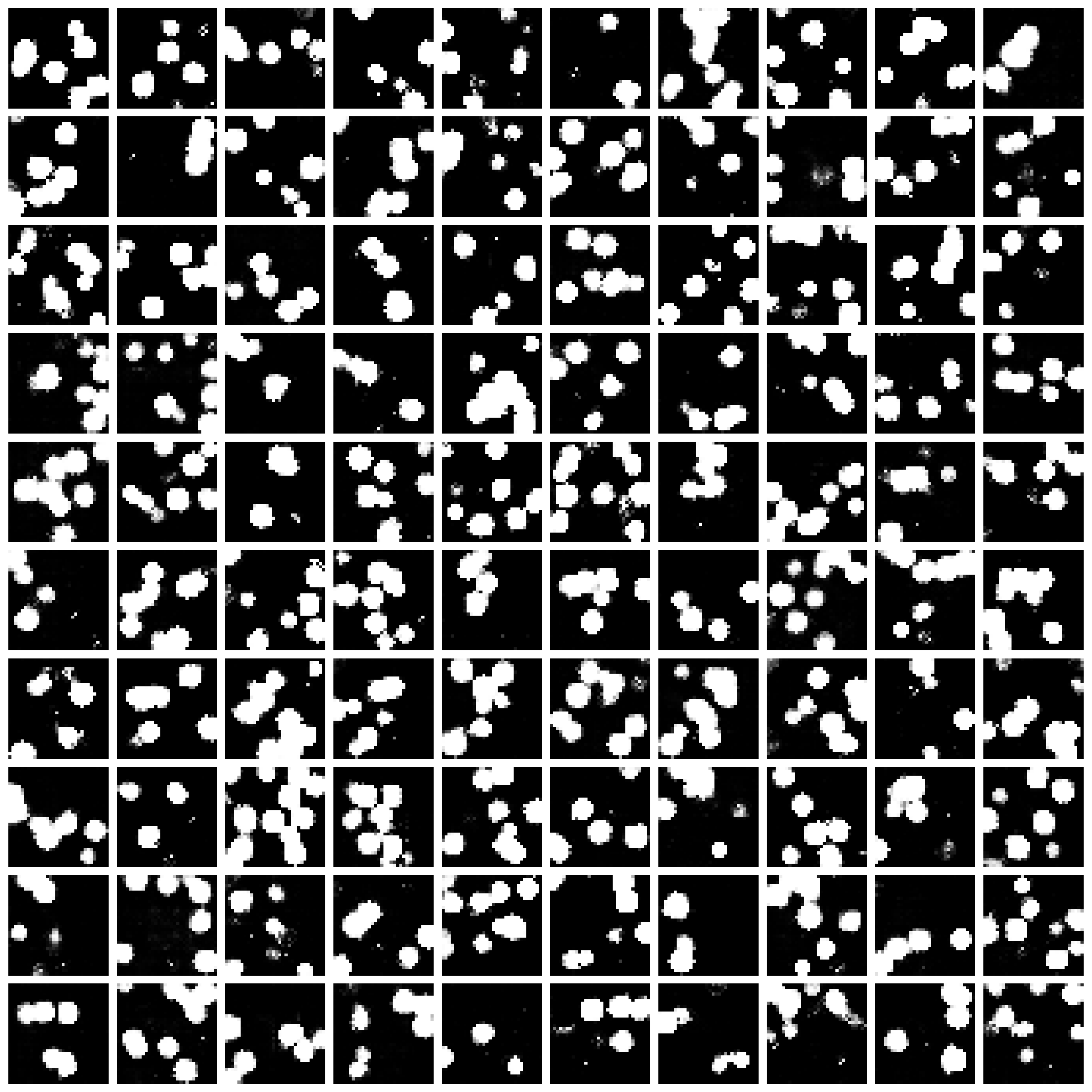}
        \caption{Synthetic}
        \label{fig:gan}
    \end{subfigure}
    \caption{\textbf{Example 1: Best model.} Spherical inclusions
        embedded in a matrix material. Center plane slices of $1 \times 10^2$
        three-dimensional microstructures. (a) Original data. (b) Synthetic
        microstructures generated by a generator with $n_f^{\mathcal{G}} = 64$
        filters trained by a discriminator with $n_f^{\mathcal{D}} = 64$
    filters. The standard deviation for $\bm{z}_{\text{pred}}$ was chosen as
    $\sigma_{\text{pred}} = 10$.
Details can be found in \autoref{sec:best}.}
    \label{fig:best}
\end{figure}

\noindent
Finally, the best choice for the discriminator and generator were chosen as
described in \autoref{sec:filter_d} and \autoref{sec:filter_g}, respectively.
The GAN was trained for $n_{\text{iter}} = 2 \times 10^6$ steps on $n_s = 1
\times 10^4$
training samples. The
resulting metrics for $n_s = 1 \times 10^3$ different samples can be seen in
\autoref{tab:best}, \autoref{fig:best2p} and \autoref{fig:best}. The relative
error of the 1-point probability function as defined in \autoref{eq:error},
which is identical to the inclusion volume fraction defined in
\autoref{eq:1point}, is $\mathcal{E}_{{p}_1} = 3.464 \times 10^{-2}$. The
relative error of the
2-point probability function defined in \autoref{eq:2point} is
$\mathcal{E}_{{p}_2} = 2.438 \times 10^{-2}$. The 2-point probability
distribution is illustrated in \autoref{fig:best2p}. The relative error of the
homogenized $\bar{\mathbb{C}}_{11}$ component as defined in
\autoref{sec:homogenization} is $\mathcal{E}_{\bar{\mathbb{C}}_{11}} = 3.907
\times 10^{-3}$. As can be seen in
\autoref{tab:best}, the error with respect to the homogenized
$\bar{\mathbb{C}}_{11}$ component is one magnitude lower than the errors in the
n-point correlation functions and way below $1\%$ relative error. This shows,
that the microstructural descriptors are not the primary drivers for the
homogenized properties of the material. In this sense, the proposed GAN offers
an advantage over descriptor driven generative approaches, as these do not enter
explicitly in the optimization formulation. Moreover, the GAN learns the optimal
descriptors during training only through the Wasserstein loss.

The results indicate, that the proposed GAN is able to generate
highly accurate synthetic microstructures in the sense of micromechanical
homogenized properties with respect to a training dataset.

\begin{table}[htb] 
    \centering 
    \begin{tabular}{@{}lll@{}} 
        \toprule 
        $\mathcal{E}_{{p}_1}$ & 
        $\mathcal{E}_{{p}_2}$ & 
        $\mathcal{E}_{\bar{\mathbb{C}}_{11}}$ \\
        \midrule 
        $3.463 \times 10^{-2}$ 
        & $2.438 \times 10^{-2}$ 
        & $3.907 \times 10^{-3}$ 
        \\
        \bottomrule 
    \end{tabular} 
    \caption{\textbf{Example 1: Best model:} Error measures for a generator with
    $n_f^{\mathcal{G}} = 64$ filters trained by a discriminator with
$n_f^{\mathcal{D}} = 64$ filters for the best models as explained in
\autoref{sec:best}.} 
    \label{tab:best} 
\end{table}

\FloatBarrier
\section{Example 2: Micro CT scan} 
\label{sec:example2} 
\begin{figure}[htb]
    \centering
    \includegraphics[width=0.5\textwidth]{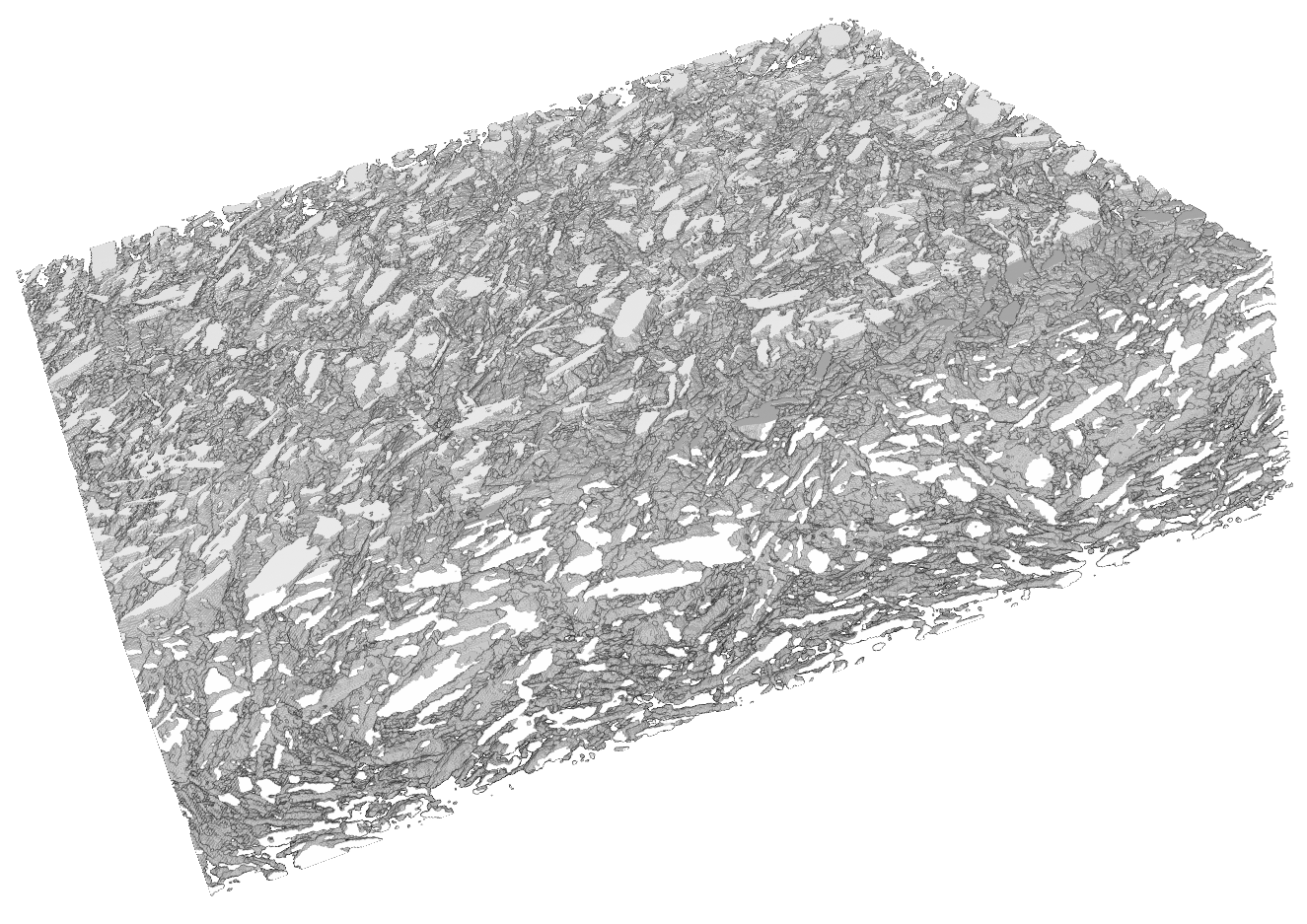}
    \caption{Micrograph of WPC as used in \autoref{sec:example2}.}
    \label{fig:micrograph}
\end{figure}

\noindent
For the second numerical experiment, a GAN is used to generate synthetic
microstructures of a real-world $\mu$CT-scan of a wood-plastic composite (WPC).
A filtered, binary micrograph with dimensions $950 \times 240 \times 850$ voxel,
which is a total of $193.8 \times 10^6$ voxels, is sliced into $1 \times 10^3$
sub-volumes with dimensions $64 \times 64 \times 64$ via Latin-Hypercube
sampling \cite{mckay2000comparison}. Every sub-volume has therefore $2.6 \times
10^5$ voxels. The same material parameters for the matrix and inclusion as in
\autoref{sec:example1} were chosen. The original $\mu$CT is illustrated in
\autoref{fig:micrograph}.

Because the image size is larger then the microstructures used in
\autoref{sec:example1}, an additional block has been included for both the
generator and the discriminator, resulting in $n_L^{\mathcal{G}} = 6$ and
$n_L^{\mathcal{D}} = 6$. For the generator, $n_f^{\mathcal{G}} = 64$ filters
in the last layer have been chosen, which results in a total of
$\bm{\theta}_{\mathcal{G}} = 21 \times 10^6$ trainable weights. Given the
number of training samples, for the discriminator, $n_f^{\mathcal{D}} = 8$
filters in the first layer have been chosen, which results in a total of
$\bm{\theta}_{\mathcal{D}} = 3.6 \times 10^6$ trainable weights. The training
was carried out for $n_{\text{iter}} = 3 \times 10^6$ steps.

\begin{figure}[htb]
    \centering
    \begin{tikzpicture}
        \begin{axis}[ 
                xlabel={correlation length},
                ylabel={${{p}_2} (\bm{x}) / p_2 (0)$},
                legend pos=outer north east,
            ] 
            \addplot[color=black, mark=x, mark options=solid] 
            coordinates {

                (1, 0.8523169)  
                (2, 0.75716454)
                (3, 0.63770425) 
                (4, 0.5496946 ) 
                (5, 0.474071  ) 
                (6, 0.41291845)
                (7, 0.3693788 ) 
                (8, 0.33568248) 
                (9, 0.3078454 ) 
                (10, 0.28684834) 
                (11, 0.27041587) 
                (12, 0.25724974)
                (13, 0.2468128 ) 
                (14, 0.23810132) 
                (15, 0.23124607) 
                (16, 0.22577819) 
                (17, 0.22129826) 
                (18, 0.21775496)
                (19, 0.21479264) 
                (20, 0.2123236 ) 
                (21 0.21026768) 
                (22, 0.20870477) 
                (23, 0.20731893) 
                (24, 0.20617722)
                (25, 0.20525946) 
                (26, 0.20454952) 
                (27, 0.20391652) 
                (28, 0.20338811) 
                (29, 0.2030021 ) 
                (30, 0.20266873)
                (31, 0.20240027)
            };
            \addlegendentry{Data}

            \addplot[color=black, dashed, mark=o, mark options=solid] 
            coordinates {
                (1, 0.7876813  )
                (2, 0.6838903  )
                (3, 0.57507163 )
                (4, 0.49549973 )
                (5, 0.43108487 )
                (6, 0.3780518  )
                (7, 0.3416278  )
                (8, 0.31303576 )
                (9, 0.29037493 )
                (10, 0.2727008  )
                (11, 0.259583   )
                (12, 0.24888492 )
                (13, 0.24095419 )
                (14, 0.2340047  )
                (15, 0.22881386 )
                (16, 0.22415356 )
                (17, 0.22070724 )
                (18, 0.21775366 )
                (19, 0.21539578 )
                (20, 0.21325214 )
                (21, 0.21169879 )
                (22, 0.21033157 )
                (23, 0.2092812  )
                (24, 0.20828362 )
                (25, 0.20761739 )
                (26, 0.20694792 )
                (27, 0.20647556 )
                (28, 0.20589203 )
                (29, 0.20561999 )
                (30, 0.20523383 )
                (31, 0.20503439 )
            };
            \addlegendentry{GAN}

        \end{axis}
    \end{tikzpicture}
    \caption{\textbf{Example 2: WPC:} Two-point correlation
    functions for the WPC and the discriminator as described in
\autoref{sec:example2}.}
    \label{fig:wpc2p} 
\end{figure}
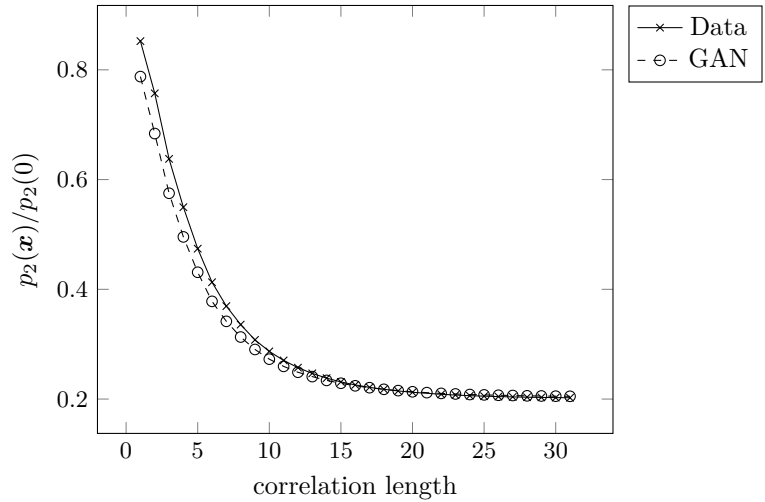

\begin{figure}[htb]
    \centering
    \begin{subfigure}[b]{0.45\textwidth}
        \centering
        \includegraphics[width=1.0\textwidth]{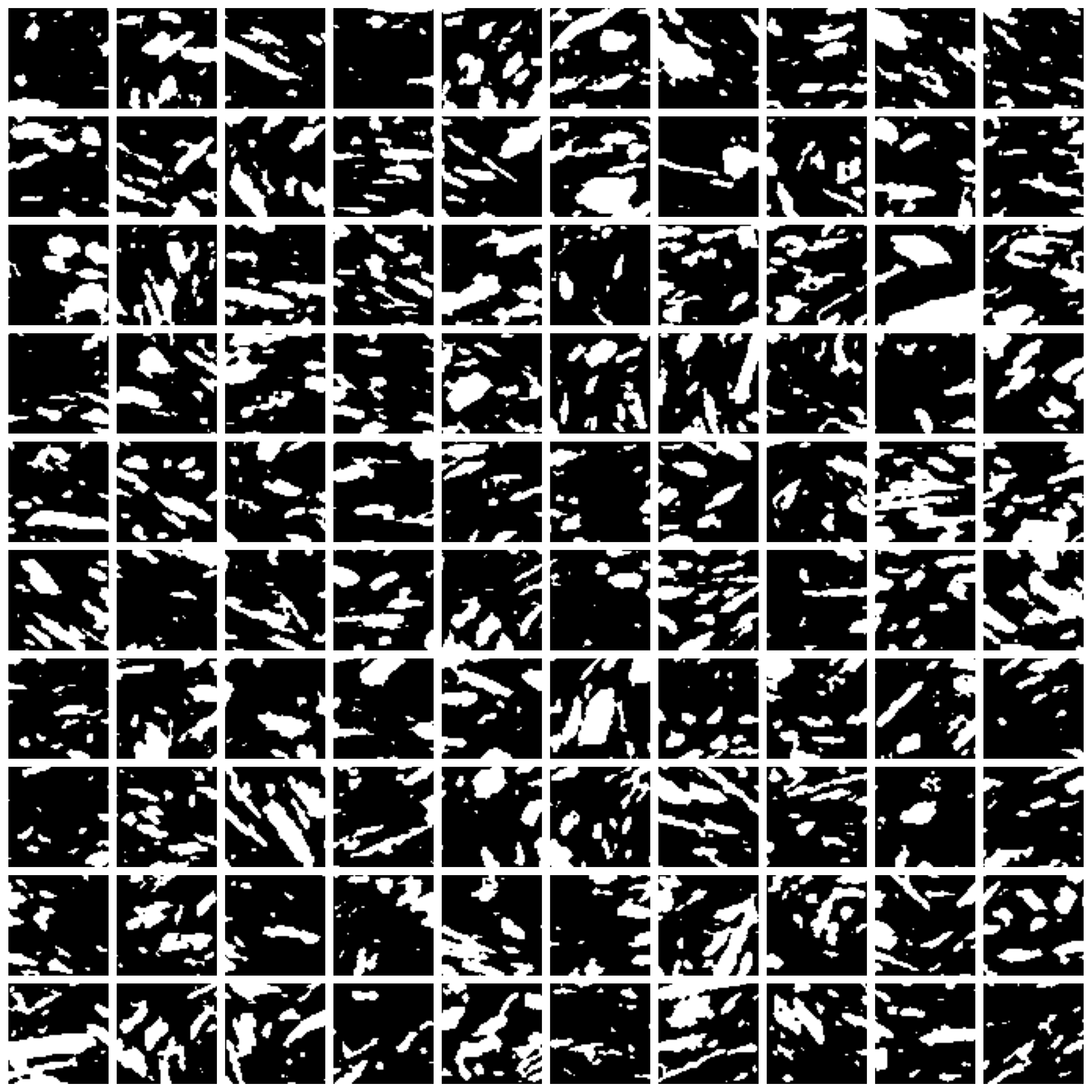}
        \caption{Real}
        \label{fig:data_wpc}
    \end{subfigure}
    \hfill
    \begin{subfigure}[b]{0.45\textwidth}
        \centering
        \includegraphics[width=1.0\textwidth]{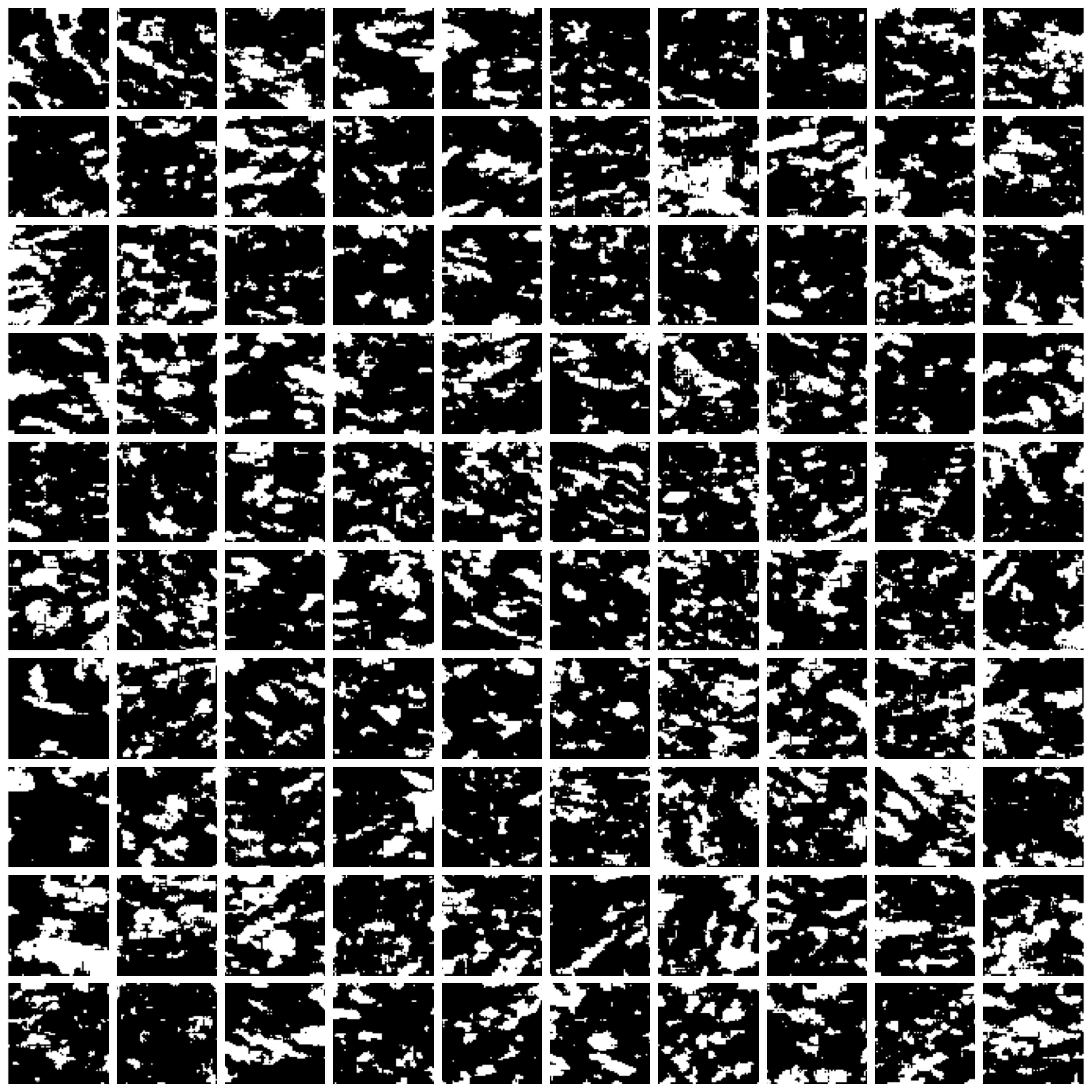}
        \caption{Synthetic}
        \label{fig:gan_wpc}
    \end{subfigure}
    \caption{\textbf{Example 2: Training microstructure.} Short wood fiber
        inclusions embedded in a polymer matrix material. Center plane slices of
        $1 \times 10^2$ three-dimensional microstructures. (a) Original data.
    (b) Synthetic microstructures generated by a GAN from
\autoref{sec:example2}.} 
\label{fig:best_wpc}
\end{figure}

The resulting metrics for $n_s = 1 \times
10^3$ different samples can be seen in \autoref{tab:wpc}. The relative error as
defined in \autoref{eq:error} of the 1-point probability function or inclusion
volume fraction defined in \autoref{eq:1point} is $\mathcal{E}_{{p}_1} =
1.018 \times 10^{-2}$. The relative error of the 2-point probability function
defined in \autoref{eq:2point} is $\mathcal{E}_{{p}_2} = 7.492 \times
10^{-2}$. The 2-point
probability distribution of both the original $\mu$CT-scan and the generator
are illustrated in \autoref{fig:wpc2p}. The relative error of the
homogenized $\bar{\mathbb{C}}_{11}$ component as defined in
\autoref{sec:homogenization} is $\mathcal{E}_{\bar{\mathbb{C}}_{11}} = 7.348 
\times 10^{-3}$. Mid-plane sections of $1 \times 10^2$ realizations of both the
sub-volumes as well as the synthetic three-dimensional images are depicted in
\autoref{fig:best_wpc}.

It can be seen from the resulting error measures as well as from the generated
images, that the proposed GAN is able to produce high quality synthetic
microstructures on the basis of real-world $\mu$CT scans. It has to be pointed
out, that only a single micrograph was used to train the GAN.

\begin{table}[htb] 
    \centering 
    \begin{tabular}{@{}lll@{}} 
        \toprule 
        $\mathcal{E}_{{p}_1}$ & 
        $\mathcal{E}_{{p}_2}$ & 
        $\mathcal{E}_{\bar{\mathbb{C}}_{11}}$ \\
        \midrule 
        $3.633 \times 10^{-2}$ 
        & $5.121 \times 10^{-2}$ 
        & $3.188 \times 10^{-2}$ 
        \\
        \bottomrule 
    \end{tabular} 
    \caption{\textbf{Example 2: WPC:} Error measures for a generator trained on
    a real $\mu$CT-Scan from \autoref{sec:example2}.} 
    \label{tab:wpc} 
\end{table}

\FloatBarrier
\section{Conclusion and outlook} 
\label{sec:end}\noindent
In the present study, a GAN for three-dimensional microstructure generation was
proposed based on a convolutional-residual discriminator and a
convolutional-residual generator with a nonlinear mapping network using a
Wasserstein loss. Several numerical examples on different microstructures
investigated the influence of hyperparameters on the synthesis quality with
respect to evaluation metrics, such as n-point correlation functions and
homogenized elasticity tensor components.

More specifically, the influence of the network topology was considered. A
growing discriminator topology as well as a constant topology was compared,
each with the same number of parameters. Both choices are present in the
literature. It was shown, that the constant topology not only was more time
consuming to train, but also suffered from slower convergence. Therefore, the
growing topology was used throughout the rest of the paper for both the
generator and discriminator. 

Next, the influence of the number of filters of the discriminator with respect
to the evaluation metrics was investigated. It was shown, that more filters lead
to lower errors, if a large number of training samples are present. For smaller
numbers of training samples, an overfitting behavior was recognized, which led
to rising errors. This is an important finding due to its profound effects on
practical considerations and has not yet been reported in the related
literature.  Generally, only a couple of hundreds of microstructures can be
sliced from a handful of $\mu$CT-scans.  Therefore, the capacity of the
discriminator has to be balanced out with regard to the number of training
samples. The effect of augmentation was not investigated, as in the specific use
case presented in
this work, the actual orientation of the microstructure is important due to its
physical response with respect to defined loading directions. If the training
microstructures are, e.g., rotated during augmentation, it is difficult
to keep track of these manipulation. 

Similar experiments were carried out for the generator, where the error went
down for larger filter numbers, at the price of slower convergence. The general
recommendation would be to use the largest generator possible having the
hardware and project time schedule in mind. 

An interesting effect with practical implications was observed at training time.
Here, larger networks showed the tendency to produce microstructures with low
variance and artifacts, if a random input vector with small standard deviation
was used. For smaller networks, this effect was reduced. Using larger standard
deviations in the input vector led to high quality, highly diverse
microstructures with the desired properties.

The last experiments dealt with a real-world WPC micrograph. From this single
micrograph, several sub-volumes were extracted. Given the limited amount of
training samples, a small discriminator was chosen. The resulting high quality
synthetic microstructures in combination with low error metrics underlined the
generative capability of the proposed GAN, which is smaller in the sense of
overall parameters as other state of the art generative networks proposed in the
literature. This enables practitioners to train the network on a single
workstation. \ah{It has to be pointed out, that the algorithm presented is able
    to extract information from a small number of $\mu$CT-scans to create
    artificial microstructures. Consequently, these microstructures only
resemble the morphology of the training sample. Effects, which lie outside the
considered training sample, cannot be captured. Here, more training data is
needed.}

Geometrical and physical inductive biases were considered but their practical
use was declined. For a brief discussion about this, see the appendices.

Future communication will deal with the comparison with the wide range of other
techniques of microstructure generation. Especially, a well established research
area is the generation of microstructures using stochastic methods, e.g.,
\cite{wang2015modelling} \cite{bortolussi2018morphological},
\cite{willot2011elastic}, \cite{abdallah2016morphological},
\cite{neumann2019stochastic}. \ah{Furthermore, the framework will be tested
against more microstructural descriptors, especially the full elasticity tensor.
Additionally, different materials, including more diverse morphologies of the
underlying microstructure, such as concrete with pores, need to be explored.
Some are currently under investigation by the authors.}

The potential of this technology in engineering applications is high. To train
e.g., a CNN for microstructure defect recognition or carrying out full field
homogenization, such as in \cite{henkes2021deep}, a large number of training
microstructures is
needed. The number of costly $\mu$CT-scans can be significantly reduced, as the
proposed GAN is able to efficiently amplify small datasets. The present work
therefore closes a gap towards an end-to-end ANN driven homogenization
framework, which is capable of predicting effective material properties from the
microscale, trained on a small number of $\mu$CT-scans. Applications of these
highly efficient approaches are real-time material property monitoring, i.e.,
structural health supervision, and multiscale simulations.
\appendix
\section{S4-equivariant CNNs}
\label{sec:group}\noindent
In the literature, geometrical constraints to ANNs, exploiting equivariance
to certain group operations such as translation and rotation, were introduced.
The idea stem from the fact, that the convolution and cross-correlation
operation in standard CNN are translational equivariant. Therefore, it is
natural to think of different symmetry group operations, which aim to improve
the accuracy and the training process. General equivariant ANNs,
especially, \textit{group-equivariant convolutional neural networks} (G-CNN),
were first introduced for in \cite{cohen2016group}. Several works in two
dimensions \cite{cohen2016steerable, cohen2018spherical, cohen2019gauge,
cohen2021equivariant} and in three dimensions \cite{worrall2018cubenet,
weiler20183d, winkels20183d, bronstein2021geometric} followed.  The idea was
adapted to equivariant GANs for two-dimensional images in \cite{karras2021alias,
dey2020group} and for three-dimensional brain image data in
\cite{dey2021generative}. In these works it was argued by the authors, that
G-CNNs can reduce the amount of training images needed to avoid overfitting.

The ordinary \textit{convolutional neural network} (CNN) using
$\bm{h}_{\star}^{(l)}$ from \autoref{eq:hidden} is translational equivariant.
This enables the ANN to learn features of the image in different locations in
the image, as opposed to the case of dense networks consisting only of dense
layers $\bm{h}_{\cdot}^{(l)}$ from \autoref{eq:hidden}, which has to learn every
feature for every location in the image.
To further include rotational equivariance, in this work a S4-equivariant CNN
was investigated, following \cite{worrall2018cubenet}. S4 is the symmetry group
of all $90 \degree$ rotations of a cube. It is
an expansion of the ordinary convolutional layer, where for each weight matrix
$\bm{W}$ a rotated copy is created. This rotation can be achieved by the group
action of S4, namely permutation of the entries of $\bm{W}$ by a precomputed
index matrix. Then, the rotated filters are shifted or convoluted over the
previous layer, such that
\begin{equation}
    \bm{h}_{\text{S4}}^{(l)} = \left\{h_{\text{S4}, \eta}^{(l)}, \; 
        \eta = 1, \ldots,
    n_{u}\right\}, \qquad h_{\text{S4}, \eta}^{(l)} = 
    \varphi^{(l)} \circ \;
    \phi^{(l)}\left(\mathbb{P}(\bm{W}^{(l)}_{\eta}) \star \bm{h}^{(l-1)}\right), 
    \label{eq:s4cnn}
\end{equation}
where $\mathbb{P}$ is a permutation matrix.
Details can be found in \cite{worrall2018cubenet}. 
In this work, the S4-convolutional layer was implemented as a custom TensorFlow
2 Keras layer to make use of GPU acceleration and parallelization. 

Whereas the resulting microstructures are qualitatively good, the utilization of
the S4 equivariant network was very challenging from a computational point of
view, as for every learned filter a number of rotated copies has to be generated
and stored. For practical applications, this becomes almost infeasible, as the
wall time between the network presented in this work versus the S4 network is in
the order of several hours to days versus several days to weeks, using a single
workstation. For every unique filter, 24 copies have to be stored. Therefore,
due to the large number of filters needed, the growing filter architecture is
infeasible, such that the less performant constant layout has to be chosen.

\begin{figure}[htb]
    \centering
    \includegraphics[width=0.5\textwidth]{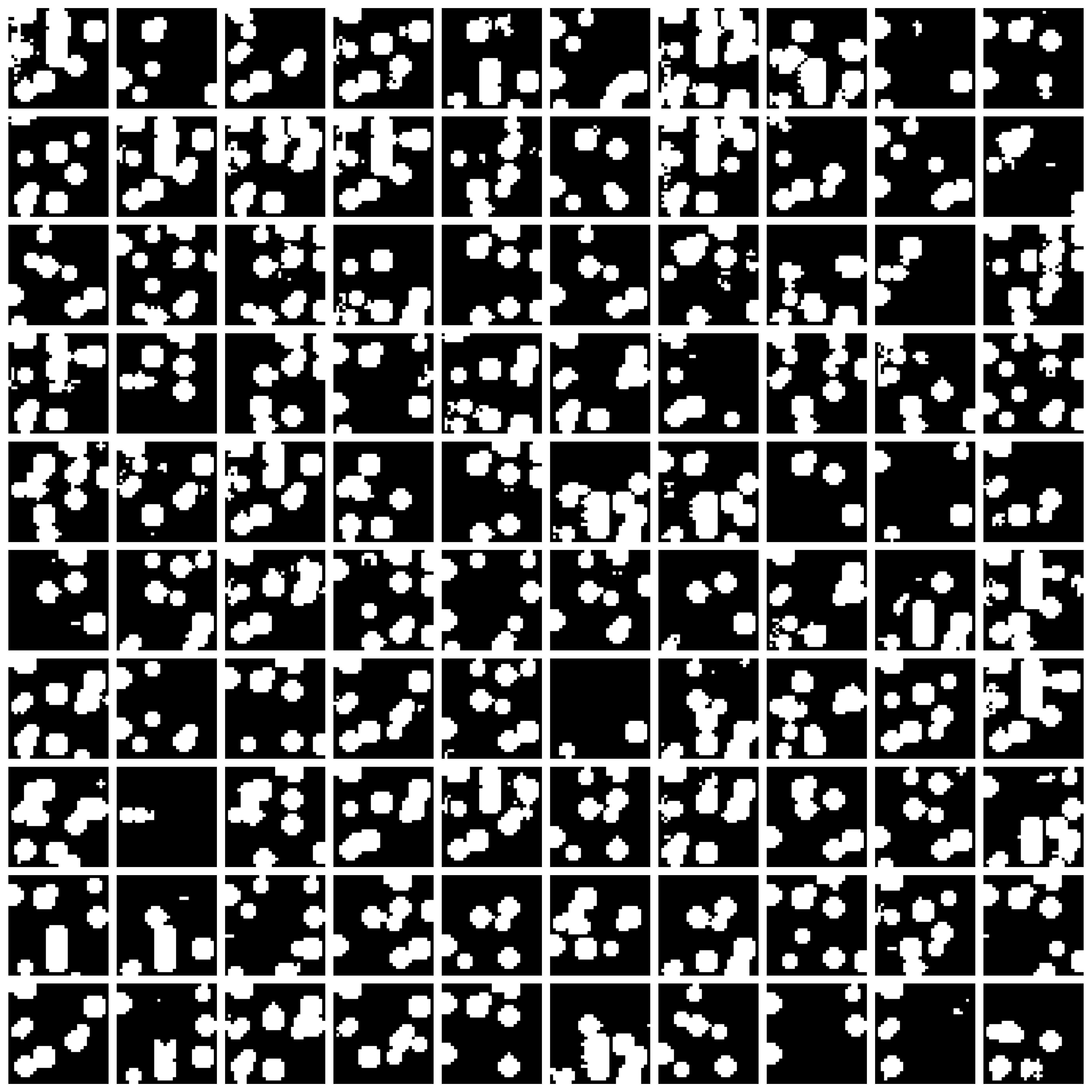}
    \caption{Center plane slices of $1 \times 10^2$ 3D RVEs. Prediction by S4
    network from \autoref{eq:s4cnn}.}
    \label{fig:microstructure_s4}
\end{figure}

Nevertheless, it is shown, that constant filter S4 discriminators provide good
results. For the spherical inclusions investigated in \autoref{sec:example1}, a
standard generator with $n_f^{\mathcal{G}} = 32$ filters was chosen. The
discriminator utilized the equivariant layers from \autoref{eq:s4cnn}. Both
networks used $n_L = 5$ blocks. Both networks were trained for $n_{\text{iter}}
= 1 \times 10^5$ iterations on $n_s = 1 \times 10^3$ training samples. Even
after this relatively small number of iterations, the error of the
$\bar{\mathbb{C}}_{11}$ component of the elasticity tensor with respect to
$10^3$ different reference samples was $\mathcal{E}_{\bar{\mathbb{C}}_{11}} =
1.5873 \times 10^{-2}$. This is a good result and the generated microstructures
are of high quality as illustrated in \autoref{fig:microstructure_s4}.
Nevertheless, the training was carried out for over one week, which makes it
currently infeasible for practical applications, even more so due to memory
limitations of current GPUs. In the future, scientist and engineers could profit
from the enhanced convergence behavior of the S4 network.

\section{Physics informed GANs}
\label{sec:PhysicGAN} \noindent
As it was shown in \cite{henkes2021physics} and \cite{wessels2020neural},
physical constraints can be introduced to ANN in the context of continuum
mechanics, which enable the network to solve the underlying partial differential
equations directly, without the need of training data. This approach is commonly
known as \textit{physics informed neural networks}.
The introduction of explicit physics into the loss function was investigated, 
e.g., in \cite{singh2018physics}. Here, the standard GAN loss as introduced in
\cite{goodfellow2014generative} was used instead of the Wasserstein loss. 

In this work, we also experimented with introducing a physical loss
$\mathcal{L}_{p_1}$, taking into account the discrepancy of the inclusion
fraction between the dataset and the generated images. Using the 1-point
correlation function $p_1$ from \autoref{eq:1point}, $\mathcal{L}_{p_1}$ is
defined as 
\begin{equation}
    \mathcal{L}_{p_1} = \mathbb{E} \left[ || p_1(\mathcal{I}_{\text{data}}) -
    p_1(\mathcal{I}_{\text{synthetic}}) ||_2^2 \right].
\end{equation}
No enhancement of the quality of the solution or the convergence behavior could
be observed. Of course, more complex physical constraints could be introduced,
e.g., Minkowski functionals or Minkowski tensors
\cite{ernesti2020characterizing}, but it is difficult to calculate these during
training time due to computational burdens. In the light of the comment
\cite{bitter}, the attempt to guide the optimization of the network by inductive
bias is ultimately hopeless, as the universal approximation capacity of the
neural network and the very general nature of the Wasserstein loss are
sufficient, given enough computational power, to capture all properties of the
microstructure at hand. This is one key advantage of the proposed GAN for
microstructure generation, as no handcrafted descriptors of the microstructures
enter the optimization process and therefore reduce human bias. Additionally,
this enables the proposed GAN to be applicable to all possible kinds of
materials, whereas descriptors, e.g., Minkowski functionals in the case of
porous materials, are often tailored towards one specific material class.

\clearpage
\section*{Acknowledgement} \noindent
We thank Neel Dey for the fruitful discussions regarding GAN and group
equivariant approaches to them.
\section*{Data availability} \noindent
The code is available on
\href{https://github.com/ahenkes1/HENKES_GAN}{https://github.com/ahenkes1/HENKES\_GAN}
and \cite{henkes_code}.
\bibliography{literature_henkes}
\bibliographystyle{elsarticle-num}
\end{document}